\documentclass[10pt,conference,twocolumn]{IEEEtran}
\usepackage{wrapfig}
\usepackage{graphicx}
\usepackage{booktabs}
\usepackage{fancyhdr}
\usepackage{amsmath}
\usepackage{amsfonts}
\usepackage{cite}
\usepackage{bm}
\usepackage{amssymb}
\usepackage{amsthm}
\usepackage{url}
\usepackage{multirow}
\usepackage{times}
\usepackage{enumitem}
\usepackage{comment}
\usepackage[linesnumbered,ruled]{algorithm2e}
\usepackage[colorlinks=true, linkcolor=black, citecolor=blue, urlcolor=blue]{hyperref}
\usepackage{siunitx}
\usepackage{mathtools}
\usepackage{tikz}
\usepackage{balance}

\makeatletter
\let\MYcaption\@makecaption
\makeatother

\usepackage[font=footnotesize]{subcaption}

\makeatletter
\let\@makecaption\MYcaption
\makeatother

\newcommand{\abs}[1]{\left|{#1}\right|}

\newcommand{\ba}{\begin{array}}
\newcommand{\ea}{\end{array}}

\renewcommand{\equiv}{\triangleq}

\newcommand{\round}[1]{\ensuremath{\lfloor#1\rceil}}

\DeclareMathOperator*{\argmin}{\arg\!\min}
\DeclareMathOperator*{\argmax}{\arg\!\max}

\let\originalleft\left
\let\originalright\right
\renewcommand{\left}{\mathopen{}\mathclose\bgroup\originalleft}
\renewcommand{\right}{\aftergroup\egroup\originalright}

\begin{document}

\title{Automotive-Radar-Based 50-cm Urban Positioning}  

\author{
\IEEEauthorblockN{Lakshay Narula, Peter A. Iannucci, Todd E. Humphreys}
\IEEEauthorblockA{\textit{Radionavigation Laboratory} \\
\textit{The University of Texas at Austin}\\
Austin, TX, USA}
}


\maketitle

\begin{abstract}
  Deployment of automated ground vehicles (AGVs) beyond the confines of sunny
  and dry climes will require sub-lane-level positioning techniques based
  on radio waves rather than near-visible-light radiation. Like human sight,
  lidar and cameras perform poorly in low-visibility conditions. This paper
  develops and demonstrates a novel technique for robust 50-cm-accurate urban
  ground positioning based on commercially-available low-cost automotive
  radars.  The technique is computationally efficient yet obtains a
  globally-optimal translation and heading solution, avoiding local minima
  caused by repeating patterns in the urban radar environment.  Performance is
  evaluated on an extensive and realistic urban data set.  Comparison against
  ground truth shows that, when coupled with stable short-term odometry, the
  technique maintains \num{95}-percentile errors below \SI{50}{\centi\meter} in
  horizontal position and \ang{1} in heading.
\end{abstract}


\newif\ifpreprint
\preprintfalse
\preprinttrue

\ifpreprint

\pagestyle{plain}
\thispagestyle{fancy}  
\fancyhf{} 
\renewcommand{\headrulewidth}{0pt}
\rfoot{\footnotesize \bf May 2020 preprint of paper accepted for publication} \lfoot{\footnotesize \bf
  Copyright \copyright~2020 by Lakshay Narula}

\else

\thispagestyle{empty}
\pagestyle{empty}

\fi

\section{Introduction}
\IEEEPARstart{D}{evelopment} of automated ground vehicles (AGVs) has spurred
research in lane-keeping assist systems, automated intersection management
\cite{fajardo2011automated}, tight-formation platooning, and cooperative
sensing \cite{choi2016mmWaveVehicular, lachapelle2020riskIcassp}, all of which
demand accurate (e.g., 50-cm at 95\%) ground vehicle positioning in an urban
environment.  But the majority of positioning techniques developed thus far
depend on lidar or cameras, which perform poorly in low-visibility conditions
such as snowy whiteout, dense fog, or heavy rain. Adoption of AGVs in many
parts of the world will require all-weather localization techniques.

Radio-wave-based sensing techniques such as radar and GNSS (global navigation
satellite system) remain operable even in extreme weather
conditions~\cite{yen2015evaluation} because their longer-wavelength
electromagnetic radiation penetrates snow, fog, and rain.
Carrier-phase-differential GNSS (CDGNSS) has been successfully applied for the
past two decades as an all-weather decimeter-accurate localization technique in
open-sky conditions.  Coupling a CDGNSS receiver with a tactical-grade inertial
sensor, as in~\cite{petovello2004benefits,scherzinger2006precise,
zhangComparisonWithTactical2006,kennedy2006architecture} delivers robust
high-accuracy positioning even during the extended signal outages common in the
urban environment, but such systems are far too expensive for widespread
deployment on AGVs.  Recent work has shown that 20-cm-accurate (95\%) CDGNSS
positioning is possible at low cost even in dense urban areas, but solution
availability remains below 90\%, with occasional long gaps between
high-accuracy solutions~\cite{humphreys2019deepUrbanIts}.  Moreover, the global
trend of increasing radio interference in the GNSS bands, whether accidental or
deliberate~\cite{humphreysGNSShandbook}, underscores the need for
GNSS-independent localization: GNSS jamming cannot be allowed to paralyze an
area's automated vehicle networks.

Clearly, there is a need for AGV localization that is low cost, accurate at the
$\approx$\num{50}-cm level, robust to low-visibility conditions, and continuously
available.  This paper is the first to establish that automotive-radar-based
localization can meet these criteria.

\begin{figure*}[htb!]
  \centering
  \begin{minipage}[b]{0.245\textwidth}
    \centering
    \includegraphics[width=\linewidth,draft=false]{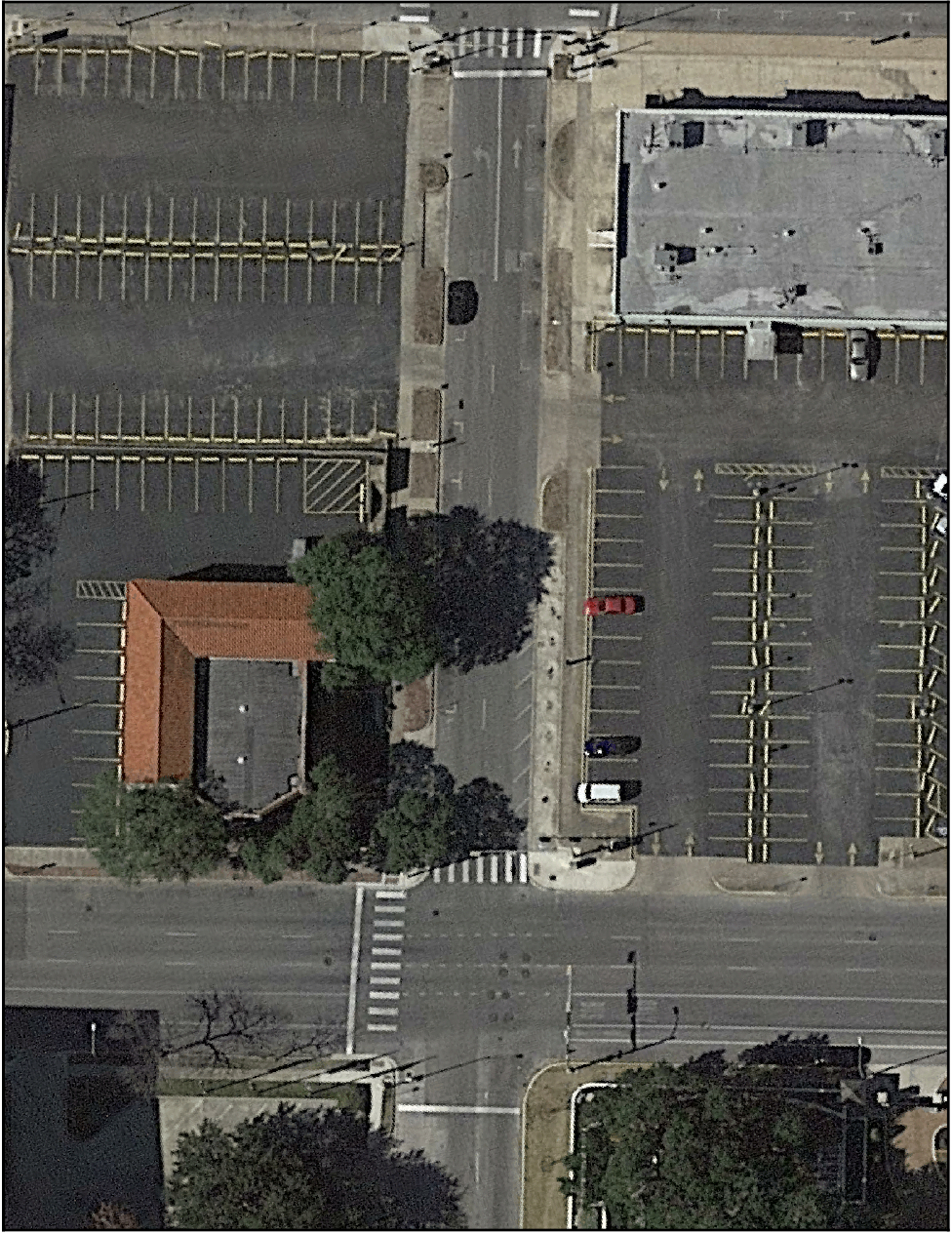}
    \subcaption{}
    \label{fig:satellite-view}
  \end{minipage}
  \begin{minipage}[b]{0.245\textwidth}
    \centering
    \includegraphics[width=\linewidth,draft=false]{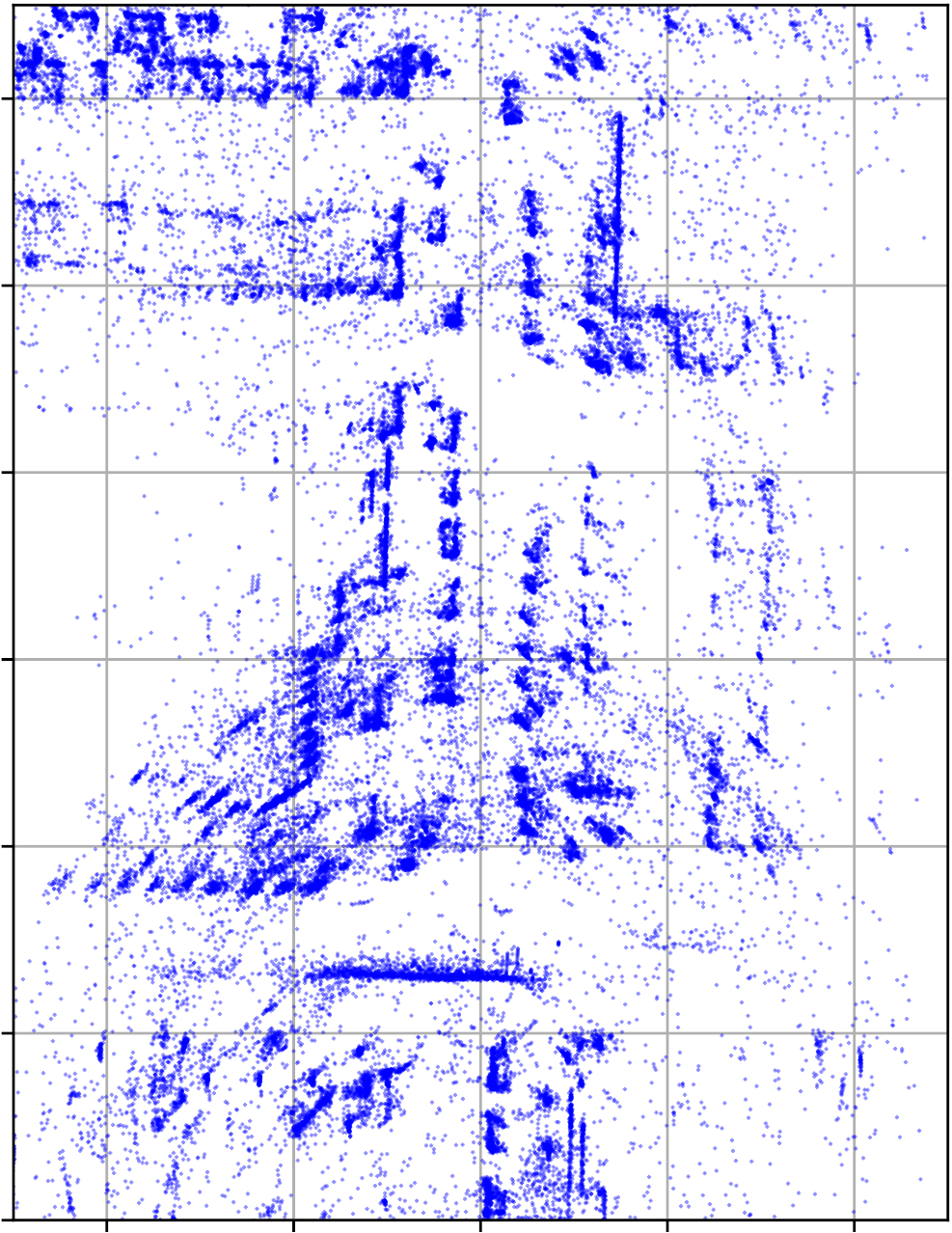}
    \subcaption{}
    \label{fig:radar-map}
  \end{minipage}
  \begin{minipage}[b]{0.245\textwidth}
    \centering
    \includegraphics[width=\linewidth,draft=false]{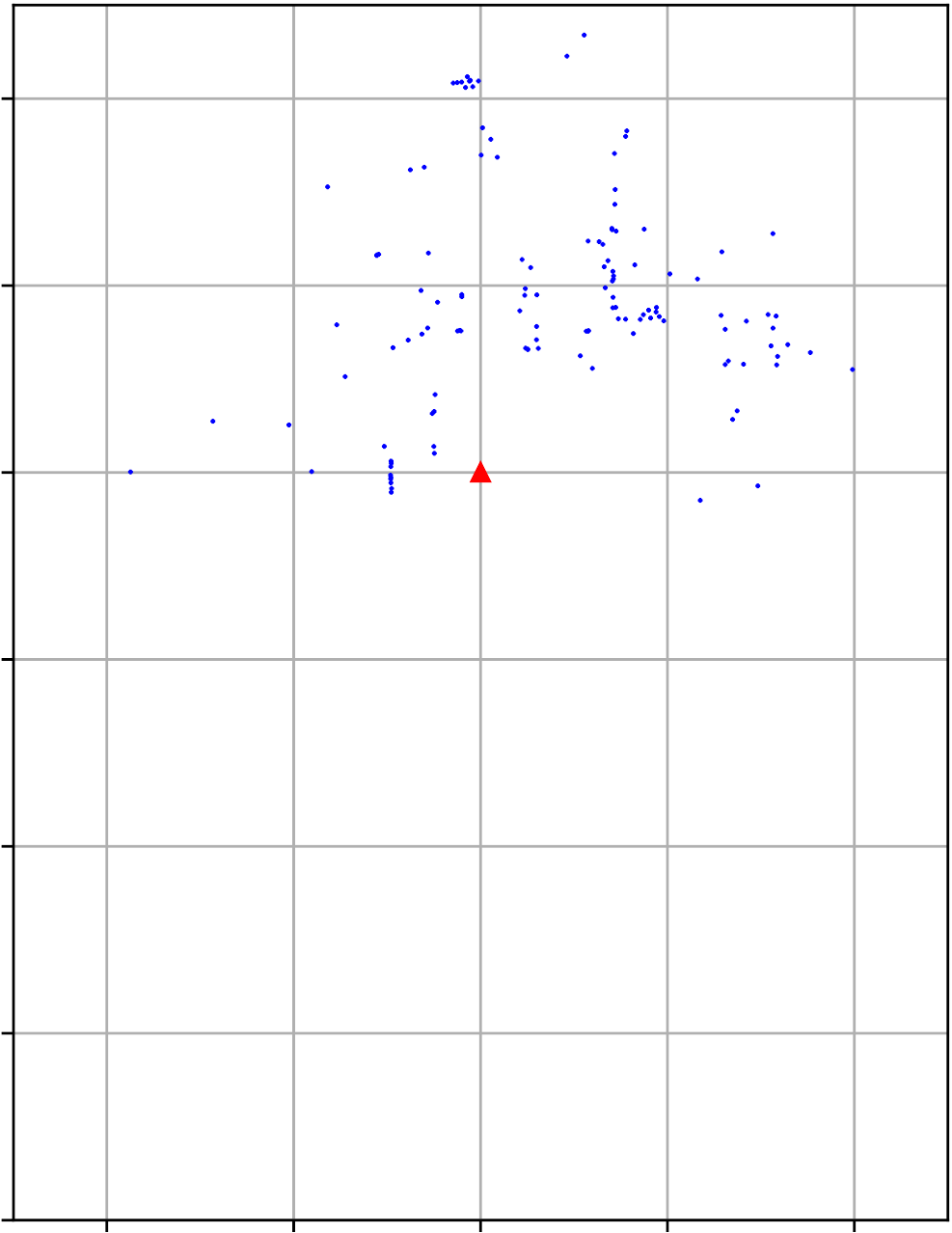}
    \subcaption{}
    \label{fig:single-scan}
  \end{minipage}
  \begin{minipage}[b]{0.245\textwidth}
    \centering
    \includegraphics[width=\linewidth,draft=false]{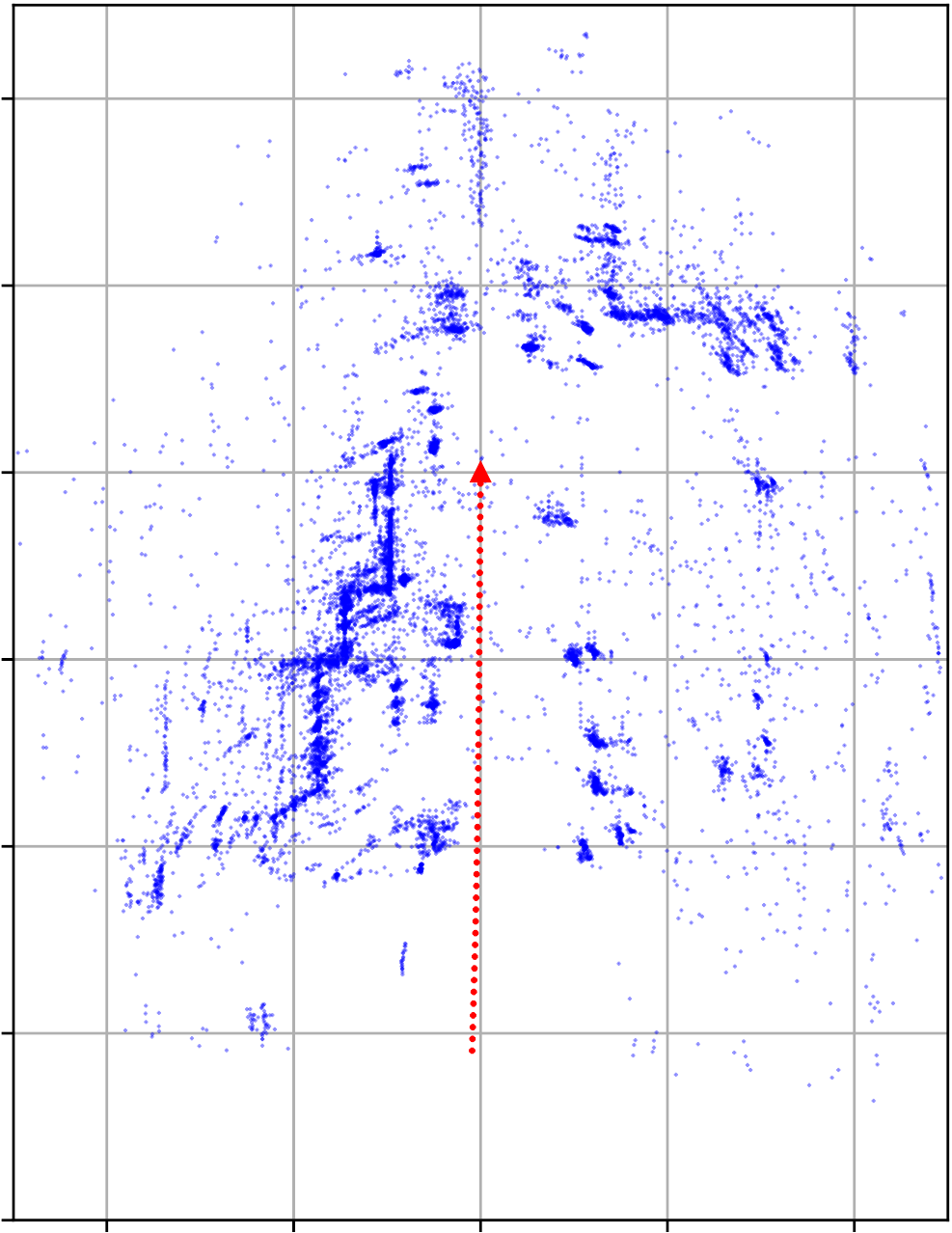}
    \subcaption{}
    \label{fig:radar-batch}
  \end{minipage}
  \caption{Panel (a) shows a satellite view of the environment being mapped
  with automotive radar. Panel (b) shows the generated radar map point cloud
  with vehicle pose obtained from a reference localization system. Note the
  repeating structure along the road side due to parked vehicles. An individual
  radar scan obtained during localization is shown in panel (c), along with the
  red triangle denoting vehicle location and heading. The scan is sparse and
  contains significant clutter, making it challenging to register to the prior
  map. Panel (d) shows a batch of radar scans during localization, with the
  red dots denoting the vehicle trajectory over the past five seconds.  The
  batch captures the underlying structure which can be registered to the
  prior map.}
  \label{fig:radar-scan-matching}
\end{figure*}

Mass-market commercialization has brought the cost of automotive radar down
enough that virtually all current production vehicles are equipped with at
least one radar unit, which serves as the primary sensor for adaptive cruise
control and automatic emergency braking.  But use of automotive radar for
localization faces the significant challenges of data sparsity and noise: an
automotive radar scan has vastly lower resolution than a camera image or a
dense lidar scan, and is subject to high rates of false detection (clutter) and
missed detection.  As such, it is nearly impossible to deduce semantic
information or to extract distinctive environmental features from an individual
radar scan.  This is clear from Fig.~\ref{fig:single-scan}, which shows a
sparse smattering of reflections from a single composite scan using three radar
units. The key to localization is to aggregate sequential scans into a batch,
as in Fig.~\ref{fig:radar-batch}, where environmental structure is clearly
evident.  Even still, the data remain so sparse that localization based on
traditional machine vision feature extraction and matching is not promising.

This paper proposes a two-step process for radar-based localization.  The first
is the mapping step: creation of a geo-referenced two-dimensional aggregated
map of all radar targets across an area of interest.  Fig.~\ref{fig:radar-map}
shows such a map, hereafter referred to as a radar map. The full radar map
used throughout this paper, of which Fig.~\ref{fig:radar-map} is a part, was
constructed with the benefit of a highly stable inertial platform so that a
trustworthy ground truth map would be available against which maps generated by
other techniques could be compared.  But an expensive inertial system or
dedicated mobile mapping vehicle is not required to create a radar map.
Instead, it can be crowd-sourced from the very user vehicles that will
ultimately exploit the map for localization.  During periods of favorable
lighting conditions and good visibility, user vehicles can exploit a
combination of low-cost CDGNSS, as in~\cite{humphreys2019deepUrbanIts}, and
GNSS-aided visual simultaneous localization and mapping, as
in~\cite{narula2018accurate}, to achieve the continuous
decimeter-and-sub-degree-accurate geo-referenced position and orientation
(pose) required to lay down an accurate radar map.  In other words, the radar
map can be \emph{created} when visibility is good and then \emph{exploited} at
any later time, such as during times of poor visibility.

Despite aggregation over multiple vehicle passes, the sparse and cluttered
nature of automotive radar data is evident from the radar map shown in
Fig.~\ref{fig:radar-map}: the generated point cloud is much less dense and has
a substantially higher fraction of spurious returns than a typical
lidar-derived point cloud, making automotive-radar-based localization a
significantly more challenging problem.

The second step of this paper's technique is the localization step.  Using some
type of all-weather odometric technique such as inertial sensing, radar
odometry, or wheel rotation and steering encoders---or a combination of
these---the estimated changes in vehicle pose over a short interval (e.g.,
\SI{5}{\second}) are used to spatially organize the multiple radar scans over
the interval and generate what is hereafter referred to as a batch of scans, or
batch for short.  Fig.~\ref{fig:radar-batch} shows a five-second batch
terminating at the same location as the individual scan in
Fig.~\ref{fig:single-scan}. In contrast to the individual scan, some
environmental structure emerges in the batch of scans, making robust
registration to the map feasible. Even so, the localization problem remains
challenging due to the dynamic radar environment: note the absence of parked
cars on the left side of the street during localization.

\textbf{Contributions.}  This paper develops a robust computationally efficient
correlation-maximization-based globally-optimal radar scan registration
algorithm that estimates a two-dimensional translational and a one-dimensional
rotational offset between a prior radar map and a batch of current scans.
Significantly, the technique can be applied to the highly sparse and cluttered
data produced by commercially-available low-cost automotive radars.
Maximization of correlation is shown to be equivalent to minimization of the
$L^2$ distance between the prior map and the batch probability hypothesis
densities.

The proposed method relies on the generation of a batch of radar scans using
odometric sensors over short time intervals. An important contribution of this
work is to analyze the robustness of the proposed method to short batch lengths
and odometric imperfections.

This paper also presents the first evaluation of low-cost automotive-grade
radar-based urban positioning on the large-scale dataset described
in~\cite{narula2020texcup}. Data from automotive sensors are collected
over two \SI{1.5}{\hour} driving sessions through the urban center of Austin,
TX on two separate days specifically chosen to provide variety in traffic and
parking patterns.  Comparison with a post-processed ground truth trajectory
shows that proposed radar-based localization algorithm has \num{95}-percentile
errors of \SI{44}{\centi\meter} in horizontal position and \ang{0.59} in
heading when using drift-free \SI{5}{\second} batches.

\textbf{Organization of the rest of this paper.} Sec.~\ref{sec:technique}
describes the theoretical underpinnings of the proposed localization technique
and explains its modeling assumptions. Sec.~\ref{sec:related-work} compares and
contrasts the proposed approach with the prior work in related
fields. Implementation details concerning computational efficiency and low-cost
automotive radar sensor modeling are presented in
Sec.~\ref{sec:implementation}. Experimental results on field data are detailed
and evaluated in Sec.~\ref{sec:results}, and Sec.~\ref{sec:conclusion} gives
concluding remarks.

\section{Localization Technique}
\label{sec:technique}

This section describes the theoretical formulation of the localization
technique adopted in this paper. It first details the statistical motivation
behind the method, and then develops an efficient approximation to the exact
solution.

\subsection{Localization using Probability Hypothesis Density}
\label{sec:phd-localization}

For the purpose of radar-based localization, an AGVs environment can be
described as a collection of arbitrarily shaped radar reflectors in a specific
spatial arrangement. Assuming sufficient temporal permanence of this
environment, radar-equipped AGVs make sample measurements of the underlying
structure over time.

\subsubsection{The Probability Hypothesis Density Function}

A probabilistic description of the radar environment is required to set up
radar-based localization as an optimization problem. This paper chooses the
probability hypothesis density (PHD)
function~\cite{mahler2003multitargetFirstOrder} representation of the radar
environment. The PHD at a given location gives the density of the expected
number of radar reflectors at that location. For a static radar environment,
the PHD $D(\bm{x})$ at a location $\bm{x} \in \mathcal{X}$ can be written as
\[
  D(\bm{x}) = I \cdot p(\bm{x})
\]
where $\mathcal{X}$ is the set of all locations in the environment, $p(\bm{x})$
is a probability density function such that $\int p(\bm{x}) {\rm d}\bm{x} = 1$,
and $I$, the intensity, is the total number of radar reflectors in the
environment. For a time-varying radar environment, both $I$ and $p(\bm{x})$ are
functions of time. For $\mathcal{A} \subset \mathcal{X}$, the expected number
of radar reflectors in $\mathcal{A}$ is given as
\[
  I_\mathcal{A} = \int_\mathcal{A} D(\bm{x}) {\rm d}\bm{x}
\]

\subsubsection{Estimating Vehicle State from PHDs}

Let $D_{\rm m}(\bm{x})$ denote the ``map'' PHD function representing the
distribution of radar reflectors in an environment, estimated as a result of
mapping with known vehicle poses. During localization, the vehicle makes a
radar scan, or a series of consecutive radar scans. A natural solution to the
vehicle localization problem may be stated as the vehicle pose which maximizes
the likelihood of the observed batch of scans, given that the scan was drawn
from $D_{\rm m}(\bm{x})$~\cite{myronenko2010point}. This maximum likelihood
estimate (MLE) has many desirable properties such as asymptotic efficiency.
However, the MLE solution is known to be sensitive to outliers that may occur
if the batch of scans was sampled from a slightly different PHD, e.g., due to
variations in the radar environment between mapping and
localization~\cite{jian2010robust}.

A more robust solution to the PHD-based localization problem may be stated as
follows.  Let $\bm{\Theta}$ denote the vector of parameters of the rigid or
non-rigid transformation $\mathcal{T}$ between the vehicle's prior belief of
its pose, and its true pose. For example, in case of a two-dimensional rigid
transformation, $\bm{\Theta} = {\left[ \Delta x, \Delta y, \Delta \phi
\right]}^\top$, where $\Delta x$ and $\Delta y$ denote a two-dimensional
position and $\Delta \phi$ denotes heading.  Also, let $D_{\rm
b}(\bm{x}^\prime)$ denote a local ``batch'' PHD function estimated from a batch
of scans during localization, defined over $\bm{x}^\prime \in \mathcal{A}
\subset \mathcal{X}$.  This PHD is represented in the coordinate system
consistent with vehicle's prior belief, such that $\bm{x}^\prime =
\mathcal{T}_{\bm{\Theta}}(\bm{x})$.  Estimating the vehicle pose during
localization is defined as estimating $\bm{\Theta}$ such that some distance
metric between the PHDs $D_{\rm m}(\bm{x})$ and $D_{\rm b}(\bm{x}^\prime)$ is
minimized.

This paper chooses the $L^2$ distance between $D_{\rm m}(\bm{x})$ and $D_{\rm
f}(\bm{x}^\prime)$ as the distance metric to be minimized. As compared to the
MLE which minimizes Kullback-Leibler divergence, $L^2$ minimization trades off
asymptotic efficiency for robustness to measurement model
inaccuracy~\cite{jian2010robust}. The $L^2$ distance $d_{L^2}(\bm{\Theta})$ to
be minimized is given as
\[
  d_{L^2}(\bm{\Theta}) = \int_\mathcal{A} {\left( D_{\rm m}(\bm{x}) - D_{\rm b}(\mathcal{T}_{\bm{\Theta}}(\bm{x})) \right)}^2 {\rm d}\bm{x}
\]

For rigid two-dimensional transformations, it can be shown as follows that
minimizing the $L^2$ distance between the PHDs is equivalent to maximization of
the cross-correlation between the PHDs.
\begin{align*}
  \widehat{\bm{\Theta}} &= \argmin_{\bm{\Theta}^\prime} \int_\mathcal{A} {\left( D_{\rm m}(\bm{x}) - D_{\rm b}(\mathcal{T}_{\bm{\Theta}^\prime}(\bm{x})) \right)}^2 {\rm d}\bm{x} \\
  &
  \begin{multlined}[b]
    = \argmin_{\bm{\Theta}^\prime} \left[ \int_\mathcal{A} D_{\rm m}^2(\bm{x}) {\rm d}\bm{x} + \int_\mathcal{A} D_{\rm b}^2(\mathcal{T}_{\bm{\Theta}^\prime}(\bm{x})) {\rm d}\bm{x} \right. \\
                                   \left. -2\int_\mathcal{A} D_{\rm m}(\bm{x}) D_{\rm b}(\mathcal{T}_{\bm{\Theta}^\prime}(\bm{x})) {\rm d}\bm{x} \right]
  \end{multlined}
\end{align*}
Note that the first term above is fixed during optimization, while the second
term is invariant under rigid transformation.  As a result, the above
optimization is equivalent to maximizing the cross-correlation:
\begin{equation}
  \widehat{\bm{\Theta}} = \argmax_{\bm{\Theta}^\prime} \int_\mathcal{A} D_{\rm m}(\bm{x}) D_{\rm b}(\mathcal{T}_{\bm{\Theta}^\prime}(\bm{x})) {\rm d}\bm{x}
  \label{eq:phd-correlation}
\end{equation}
For differentiable $D_{\rm m}$ and $D_{\rm b}$, the above optimization can be
solved with gradient-based methods. However, as detailed in
Sec.~\ref{sec:related-work}, the cross-correlation maximization problem in the
urban AGV environment may have locally optimal solutions in the vicinity of the
global minimum due to repetitive structure of radar reflectors. In applications
with high integrity requirements, a search for the globally optimal solution is
necessary. This paper notes that if the PHDs in (\ref{eq:phd-correlation}) were
to be discretized in $\bm{x}$, then the cross-correlation values can be
evaluated exhaustively with computationally efficient techniques. Let
$\bm{x}_{pq}$ denote the location at the $(p,q)$ translational offset in
discretized $\mathcal{A}$. Then
\begin{equation}
  \widehat{\bm{\Theta}} = \argmax_{\bm{\Theta}^\prime} \sum_{p=0}^{P-1}
  \sum_{q=0}^{Q-1} D_{\rm m}(\bm{x}_{pq}) D_{\rm b}(\round{\mathcal{T}_{\bm{\Theta}^\prime}(\bm{x}_{pq})})
  \label{eq:phd-discrete-correlation}
\end{equation}
where $\round{.}$ denotes the nearest grid point in the discretized space.

The technique developed above relies on the PHDs $D_{\rm m}$ and $D_{\rm b}$.
The next subsections detail the recipe for estimating these PHDs from the radar
observations.

\subsection{Estimating the map PHD from measurements}
\label{sec:phd-approximation}

This section addresses the procedure to estimate the map PHD $D_{\rm
m}(\bm{x})$ from radar measurements. This paper works with an occupancy grid
map (OGM) approximation to the continuous PHD function.
In~\cite{erdinc2009bin}, it has been shown that the PHD representation is a
limiting case of the OGM as the grid cell size becomes vanishingly small.
Intuitively, let $c_{pq}$ denote the grid cell region with center
$\bm{x}_{pq}$, and let $\delta c_{pq}$ denote the area of this grid cell, which
is small enough such that no more than one reflector may be found in any cell.
Let $p_{pq} (O)$ denote the occupancy probability of $c_{pq}$, and let
$\mathcal{A}$ be defined as the region formed by the union of all $c_{pq}$
whose centers $\bm{x}_{pq}$ fall within $\mathcal{A}$. Then, the expected
number of radar reflectors $\mathbb{E} \left[ \abs{\mathcal{A}} \right]$ in
$\mathcal{A}$ is given by
\begin{align*}
  \mathbb{E} \left[ | \mathcal{A} | \right] = \sum_{c_{pq} \in \mathcal{A}} p_{pq}(O)
  &= \sum_{c_{pq} \in \mathcal{A}} \frac{p_{pq}(O)}{\delta c_{pq}} \delta c_{pq} \\
  &\triangleq \sum_{c_{pq} \in \mathcal{A}} \bar{D}(\bm{x}_{pq}) \delta c_{pq} \\
  &= \int_\mathcal{A} \bar{D}(\bm{x}_{pq}) {\rm d}\bm{x}, \quad {\rm as}~\lim_{\delta c_{pq} \rightarrow 0}
\end{align*}
where $\bar{D}(\bm{x}_{pq}) \equiv \frac{p_{pq}(O)}{\delta c_{pq}}$ can be
considered to be an approximation of the PHD $D(\bm{x})$ for $\bm{x} \in
c_{pq}$ since its integral over $\mathcal{A}$ is equal to the expected number
of reflectors in $\mathcal{A}$.

The advantage of working with an OGM approximation of the PHD is two-fold:
first, since the OGM does not attempt to model individual objects, it is
straightforward to represent arbitrarily-shaped objects, and second, in
contrast to the ``point target'' measurement model assumption in standard PHD
filtering, the OGM can straightforwardly model occlusions due to extended
objects.

At this point, the task of estimating $D_{\rm m}(\bm{x})$ has been reduced to
estimating the occupancy probability of each grid cell in discretized
$\mathcal{A}$. Each grid cell $c_{pq}$ takes up one of two states: occupied
($O$) or free ($F$). Based on the radar measurement $\bm{z}_k$ at each time
$k$, the Bernoulli probability distribution of such binary state cells may be
recursively updated with the binary Bayes filter. In particular, let
$\bm{z}_{1:k}$ denote all radar measurements made up to time $k$, and let
\begin{equation}
  l_{pq}^k(O) \equiv \log{\frac{p_{pq}(O~|~\bm{z}_{1:k})}{1 - p_{pq}(O~|~\bm{z}_{1:k})}}
  \label{eq:log-odds}
\end{equation}
denote the \emph{log odds ratio} of $c_{pq}$ being in state $O$. Also define
$l_{pq}^0(O)$ as
\[
  l_{pq}^0(O) \equiv \log{\frac{p_{pq}(O)}{1 - p_{pq}(O)}}
\]
with $p_{pq}(O)$ being the prior belief on the occupancy state of $c_{pq}$
before any measurements are made. With these definitions, the binary Bayes
filter update is given by~\cite{thrun2005probabilistic}
\begin{equation}
  l_{pq}^k(O) = \log{\frac{p_{pq}(O~|~\bm{z}_k)}{1 - p_{pq}(O~|~\bm{z}_k)}} - l_{pq}^0(O) + l_{pq}^{k-1}(O)
  \label{eq:binary-bayes}
\end{equation}
where $p_{pq}(O~|~\bm{z}_k)$ is known as the \emph{inverse} sensor model:
it describes the probability of $c_{pq}$ being in state $O$, given only the
latest radar scan $\bm{z}_k$.

The required occupancy probability $p_{pq}(O~|~\bm{z}_{1:k})$ is easy to
compute from the log odds ratio in (\ref{eq:log-odds}). Observe that the
inverse sensor model $p_{pq}(O~|~\bm{z}_k)$, in addition to the prior occupancy
belief $p_{pq}(O)$, completely describes the procedure for estimating the OGM
from radar measurements, and hence approximating the PHD. Adapting
$p_{pq}(O~|~\bm{z}_k)$ to the characteristics of the automotive radar sensors,
however, is not straightforward, and will be the topic of
Sec.~\ref{sec:inverse-sensor-model}.

\subsection{Estimating the batch PHD from measurements}

The procedure for generating an approximation to $D_{\rm b}(\bm{x}^\prime)$
from a batch of radar measurements is identical to the procedure for generating
$D_{\rm m}(\bm{x})$ from mapping vehicle data, except that precise, absolute
location and orientation data is not available during localization.  Instead,
inertial odometry is used to estimate the relative locations and orientations
of each radar scan in the batch, and the scans are transformed into a common
coordinate frame before updating the occupancy state of grid cells.

Once the map and batch PHDs have been approximated from radar measurements, the
correlation-maximization technique developed in Sec.~\ref{sec:phd-localization}
can be applied to find a solution to the localization problem.

\section{Related Work}
\label{sec:related-work}

This section reviews various techniques which may be applicable to the
radar-based mapping and localization problem. These include work on point cloud
alignment and image registration techniques, occupancy grid-based mapping and
localization, and random-finite-set-based mapping and localization.

\textbf{Related work in point cloud alignment.} A radar-based map can have many
different representations.  One obvious representation is to store all the
radar measurements as a point cloud.  Fig.~\ref{fig:radar-map} is an example of
this representation.  Localization within this map can be performed with point
cloud registration techniques like the iterative closest point (ICP) algorithm.
ICP is known to converge to local minima which may occur due to outlying points
that do not have correspondences in the two point clouds being aligned. A
number of variations and generalizations of ICP robust to such outliers have
been proposed in the literature~\cite{chetverikov2002trimmed, ward2016vehicle,
tsin2004correlation, jian2010robust, myronenko2010point, gao2019filterreg}.

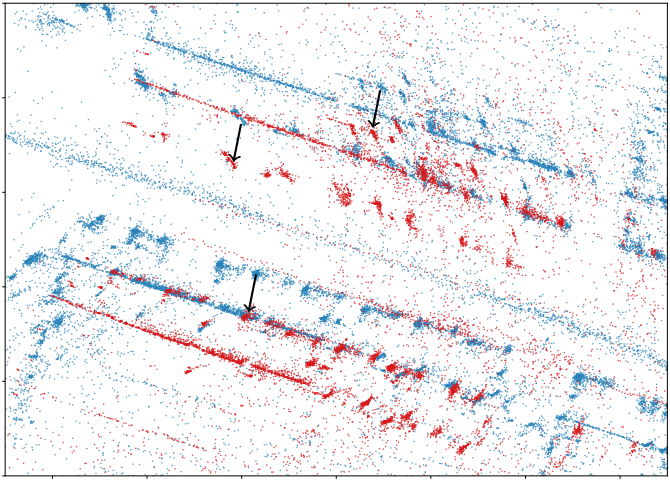
\begin{figure}[ht]
  \centering
  \begin{tikzpicture}
    \node[inner sep=0pt] at (0,0)
    {\includegraphics[width=\linewidth,draft=false] {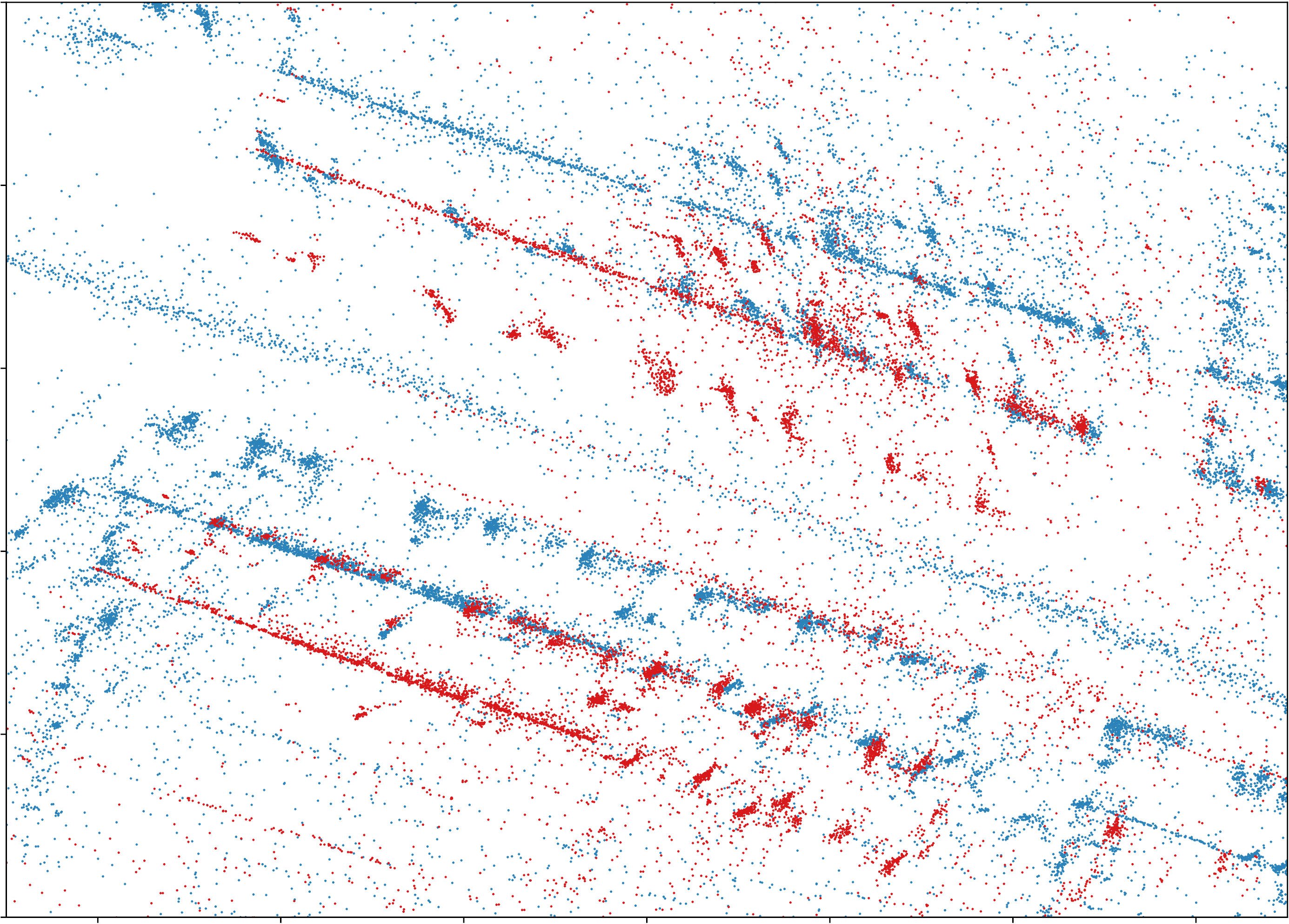}};
    \draw[->, thick] (-1.25,1.55) -- (-1.35,1.05);
    \draw[->, thick] (0.6,2.0) -- (0.5,1.5);
    \draw[->, thick] (-1.05,-0.45) -- (-1.15,-0.95);
  \end{tikzpicture}
  \caption{The radar map point cloud is shown in blue and a \SI{5}{\second}
  batch of radar scans is shown in red. Note that the map point cloud has four
  dominant parallel features: a row of parked cars flanked by a building front
  on each side of the street. The two point clouds are aligned with the CPD
  algorithm until convergence.  Note the offset of the two point clouds due to
  convergence to a local cost minimum where two of the four parallel features
  line up correctly. Black arrows show a few of the matching point features
  between the map and the batch.}
  \label{fig:local-convergence}
\end{figure}

However, automotive radar data in urban areas exhibit another source of
incorrect-but-plausible registration solutions which are not addressed in the
above literature---repetitive structure, e.g., due to a series of parked cars,
may result in multiple locally-optimal solutions within
\SIrange[range-phrase=--,range-units=single]{2}{3}{\meter} of the
globally-optimal solution. Gradient-based techniques which iteratively estimate
correspondences based on the distance between pairs of points are susceptible
to converge to such locally-optimal solutions. To demonstrate this phenomenon,
the coherent point drift algorithm (CPD)~\cite{myronenko2010point}, a robust
generalization of ICP, is applied to align the map and batch point cloud
data collected as part of this work.  Fig.~\ref{fig:local-convergence} shows an
example of incorrect convergence for CPD. Observe that the map point cloud in
Fig.~\ref{fig:local-convergence}, shown in blue, has four dominant parallel
features: a row of parked cars flanked by a building front on each side of the
street. In this scenario, when aligning a \SI{5}{\second} batch of radar scans
shown in red, the CPD algorithm converges to a locally-optimal solution where
two of the four parallel features line up correctly. This solution is about
\SI{4}{\meter} south of the globally-optimal solution.

Since a highly-accurate initial estimate of the vehicle position may not always
be available, accurate and robust radar-based urban positioning demands an
exhaustive search over the initial search region. One way to achieve
globally-optimal point cloud registration is to perform global point
correspondence based on distinctive feature descriptors instead of choosing the
closest point~\cite{cen2019radar}.  However, extraction and matching of such
features from cluttered automotive radar data is likely to be unreliable. In
view of the above limitations, a correlation-maximization-based
globally-optimal solution has been developed in this paper.

A few of the above point cloud alignment algorithms have been applied to
automotive radar data in the literature~\cite{ward2016vehicle, holder2019real}.
The technique in~\cite{ward2016vehicle} is only evaluated on a \SI{5}{\minute}
dataset, while~\cite{holder2019real} performs poorly on datasets larger than
\SI{1}{\kilo\meter}. The globally-optimal technique in~\cite{cen2019radar}
performs admirably on large and challenging datasets, but uses a more
sophisticated radar unit than is expected to be available on an AGV. Other
point cloud alignment techniques listed above have not been tested with radar
data.

\textbf{Related work in image registration and occupancy grid techniques.}
Occupancy grid mapping and localization techniques have been traditionally
applied for lidar-based systems, and recent work in~\cite{schuster2016landmark,
schoen2017real} has explored similar techniques with automotive radar data. In
contrast to batch-based localization described in this paper,
both~\cite{schuster2016landmark} and~\cite{schoen2017real} perform
particle-filter based localization with individual scans, as is typical for
lidar-based systems. These methods were only evaluated on small-scale datasets
collected in a parking lot, and even so, the reported localization accuracy was
not sufficient for lane-level positioning.

Occupancy grid maps are similar to camera-based top-down images, and thus may
be aligned with image registration techniques, that may be
visual-descriptor-based~\cite{callmer2011radar} or
correlation-based~\cite{yoneda2018vehicle}. Reliable extraction and matching of
visual features, e.g., SIFT, is significantly more challenging with automotive
radar data. Correlation-based registration is more robust, and forms the basis
of the method developed in this paper. A similar idea has previously been
proposed in~\cite{yoneda2018vehicle}. In comparison to this
paper,~\cite{yoneda2018vehicle} provides no probabilistic justification for the
approach, assumes perfect odometry information during generation of batches,
does not propose a computationally-efficient solution, and only estimates a
two-dimensional translational offset with heading assumed to be perfectly
known. Each of these extensions in the current paper is nontrivial.

\textbf{Related work in random finite set mapping and localization.} As
mentioned in Sec.~\ref{sec:technique}, the occupancy grid is an approximation
to the PHD function: a concept first introduced in the random finite set (RFS)
based target tracking literature. Unsurprisingly, PHD- and RFS-based mapping
and localization have been previously studied in~\cite{mullane2011random,
deusch2015labeled, stubler2017continuously}. In contrast to the grid-based
approximate method developed in this paper, techniques
in~\cite{mullane2011random, deusch2015labeled, stubler2017continuously} make
the point target assumption where no target may generate more than one
measurement in a single scan, and no target may occlude another target.
However, in reality, planar and extended targets such as walls and building
fronts are commonplace in the urban AGV environment. Mapping of ellipsoidal
extended targets has recently been proposed in~\cite{fatemi2017poisson}, but
occlusions are not modeled and only simulation results are presented.

\section{Implementation}
\label{sec:implementation}

Theoretical formulation of the localization technique applied in this paper was
described in Sec.~\ref{sec:technique}. This section presents some of the
implementation details necessary for a computationally efficient
globally-optimal solution, and describes the challenge of choosing an inverse
sensor model for localization based on automotive radar sensors.

\subsection{Efficient Global Optimization}

As outlined in Sec.~\ref{sec:related-work}, repetitive patterns in the urban
AGV radar environment give rise to locally-optimal solutions to the radar-based
localization problem in the vicinity of the globally-optimal
solution. Accordingly, localization problems with tight integrity requirements
demand application of techniques to find the global minimum of
(\ref{eq:phd-correlation}).  This section notes that efficient globally-optimal
procedures exist for maximizing the discretized PHD correlation as defined in
(\ref{eq:phd-discrete-correlation}), and outlines two optimizations which
further reduce the computational complexity of the problem.

\subsubsection{FFT-based Cross-Correlation}

For two-dimensional vehicle state estimation with perfect batch odometry,
the cross-correlation in (\ref{eq:phd-discrete-correlation}) is to be maximized
over the three parameters of two-dimensional rigid transformation ${ \left[
  \Delta x, \Delta y, \Delta \phi \right] }^\top$.

For a given value of $\Delta \phi$, the cross-correlation can be maximized
efficiently over $\Delta \bm{t} = { \left[ \Delta x, \Delta y \right] }^\top$
with FFT-based cross-correlation. The size of the discretized map and batch
PHDs to be correlated, denoted $P \times Q$ in
(\ref{eq:phd-discrete-correlation}), is limited by the area scanned by the
radar over a batch. Without loss of generality, let $P=Q=n$.  Due to the
convolution property of the FFT, the circular cross-correlation between $n
\times n$ matrices $D_{\rm m}(\bm{x})$ and $D_{\rm
b}(\mathcal{T}_{\bm{\Theta}}(\bm{x}))$ can be computed as
\[
  D_{\rm m} \ast D_{\rm b} = \mathcal{F}^{-1} \{ \mathcal{F} \{ D_{\rm m}(\bm{x}) \} \circ \mathcal{F} \{ D_{\rm b}( - \mathcal{T}_{\bm{\Theta}}(\bm{x})) \} \}
\]
where $\mathcal{F}$ denotes the FFT operator and $\circ$ the denotes
element-wise multiplication operator. To compute the required linear
cross-correlation, however, both $D_{\rm m}$ and $D_{\rm b}$ must be padded
with $n/2$ zeros on each side along each dimension, leading to matrices
$\check{D}_{\rm m}$ and $\check{D}_{\rm b}$ of size $2n \times 2n$.  Then, the
linear cross-correlation is
\begin{equation}
  D_{\rm m} \star D_{\rm b} = \mathcal{F}^{-1} \{ \mathcal{F} \{ \check{D}_{\rm m}(\bm{x}) \} \circ \mathcal{F} \{ \check{D}_{\rm b}( - \mathcal{T}_{\bm{\Theta}}(\bm{x})) \} \}
  \label{eq:linear-cross-correlation}
\end{equation}

The two FFTs and one IFFT in (\ref{eq:linear-cross-correlation}) each have a
computational complexity
\[
  k(2n)^2 \log{(2n)^2} \approx 8 k n^2 \log{n}
\]
where $k$ is a constant factor dependent on the FFT implementation. If the
number of rotations to be examined are $m$, this leads to a total complexity of
\begin{equation}
  3m \times 8 k n^2 \log{n} = 24 k m n^2 \log{n}
  \label{eq:basic-algo}
\end{equation}

One observation here is that $\mathcal{F} \{\check{D}_{\rm m}(\bm{x})\}$ for
the map is independent of $\Delta \phi$, and so must only be computed once.
This reduces the overall complexity to $8 k (2m + 1) n^2 \log{n}$.

\subsubsection{Minimal Padding for Desired Linear Cross-Correlation}

Typically, the size of the map and batch PHDs to be correlated, given by $P
\times Q$, is much larger than the translational offset search space due to
initial uncertainty in the vehicle position. In other words, the admissible
values of $\Delta \bm{t}$ lie within a small fraction of the space scanned in
the radar batch. Accordingly, the optimization method only requires
the linear cross-correlation values within this admissible region. If $n_l$
denotes the size of the translational search space in discretized PHD
coordinates, then $D_{\rm m}$ and $D_{\rm b}$ need only be padded with $n_l/2$
zeros on each side along each dimension, leading to matrices $\check{D}_{\rm
m}$ and $\check{D}_{\rm b}$ of size $(n + n_l) \times (n + n_l)$. With minimal
padding, the overall complexity of FFT-based correlation maximization now
reduces to
\[
  2k (2m + 1) (n + n_l)^2 \log{(n + n_l)} \approx 2k (2m + 1) n^2 \log{n}
\]
where the approximation holds if $n_l \ll n$.

\subsubsection{The FFT Rotation Theorem}

Observe from (\ref{eq:linear-cross-correlation}) that the method re-computes
the FFT after every rotation of the PHD $D_{\rm b}$. The FFT rotation
theorem~\cite{reddy1996fft} states that a coordinate rotation in the spatial
domain leads to the same coordinate rotation in the frequency domain, that is,
if $R_{\Delta \phi}$ represents the rotation matrix which operates on the PHD,
then
\[
  \mathcal{F}\{ R_{\Delta \phi} \cdot D_{\rm b}(\bm{x}) \} = R_{\Delta \phi} \cdot \mathcal{F}\{ D_{\rm b}(\bm{x}) \}
\]
This implies that instead of performing $m$ FFTs on rotated replicas of $D_{\rm
b}$, a single FFT may be performed followed by $m$ coordinate rotations. It
must be noted that rotation of $\mathcal{F}\{ D_{\rm b}(\bm{x}) \}$ could
involve interpolation of values to non-integer indices, which may offset
the computational advantage of this method. Experiments conducted as part of
this paper suggest that nearest-neighbor interpolation, i.e., assigning value
from the nearest integer index (complexity $\mathcal{O}(n^2)$), has no
discernible adverse effect on the performance of the algorithm.

With application of the FFT rotation theorem, the overall complexity is reduced
to
\[
  2k (m + 2) (n + n_l)^2 \log{(n + n_l)} + m (n + n_l)^2 \approx 2 k m n^2 \log{n}
\]
which is a factor of \num{12} faster than the basic implementation in
(\ref{eq:basic-algo}).

Algorithm~\ref{alg:correlation-search} provides the pseudocode for the
optimized FFT-based correlation maximization algorithm. For each epoch, the
algorithm is provided the prior map point cloud in the true world frame,
denoted $\bm{p}_{\rm m}^W$ and a batch of $k$ radar scans in the body frame,
denoted $\bm{p}_{{\rm b, } 1:k}^B$. An initial guess for vehicle position and
heading trajectories $\bm{t}_{1:k}^{V}$ and $\phi_{1:k}^{V}$ is provided in a
frame $V$ which is offset from the $W$ frame by a rigid two-dimensional
transform parameterized by $\bm{\Theta} = { \left[ \Delta x, \Delta y, \Delta
\phi \right] }^\top$. The initial guess uncertainties $\sigma_t$ and
$\sigma_\phi$, and the desired discretization steps $\delta t$ and $\delta
\phi$ are also provided. The algorithm must estimate the offset transformation
$\widehat{\bm{\Theta}}$ between $W$ and $V$.

The \texttt{toOGM} routine converts the provided point cloud to an occupancy
grid with the desired grid cell size, according to the procedure described in
Sec.~\ref{sec:phd-approximation}. The \texttt{pad} routine pads the provided
array with the desired number of zeros along each dimension on both ends. The
three-dimensional matrix $\mathsf{R}$ holds the linear cross-correlation
outputs.

\begin{algorithm}
  \footnotesize
  \SetKwInOut{Input}{Input}
  \SetKwInOut{Output}{Output}
  \Input{$\bm{p}_{\rm m}^W$, $\bm{p}_{{\rm b, } 1:k}^B$, $\bm{t}_{1:k}^{V}$, $\phi_{1:k}^{V}$, $\sigma_t$, $\sigma_\phi$, $\delta t$, $\delta \phi$}
  \vspace{0.25em}
  \Output{$\widehat{\bm{\Theta}}$}

  \vspace{1em}

  $D_{\rm m} = \texttt{toOGM} \left( \bm{p}_{\rm m}^W - \bm{t}_k^{V}, \delta t \right)$

  $\tilde{D}_{\rm m} = \texttt{FFT2} \left( \texttt{pad} \left( D_{\rm m}, 3\sigma_t \right) \right)$

  \vspace{1em}

  $\bm{p}_{{\rm b, } 1:k}^{V} = R \left( \phi_{1:k}^{V} \right) \bm{p}_{{\rm b, } 1:k}^B + \bm{t}_{1:k}^{V}$

  $D_{\rm b} = \texttt{toOGM} \left( \bm{p}_{{\rm b, } 1:k}^{V} - \bm{t}_k^{V}, \delta t \right)$

  $\tilde{D}_{\rm b} = \texttt{FFT2} \left( \texttt{pad} \left( D_{\rm b}, 3\sigma_t \right) \right)$

  \vspace{1em}

  $n = 3\sigma_t / \delta t$

  $m = 3\sigma_\phi / \delta \phi$

  \For{i = -m:m}{ 
    $\Delta \phi = i \delta \phi$

    $\tilde{D}_{\rm b}^{\Delta \phi} = \texttt{rotate2} \left( \tilde{D}_{\rm b}, \Delta \phi \right)$

    $\mathsf{R}[i,:,:] = \texttt{IFFT2} \left( \tilde{D}_{\rm m} \circ \texttt{conj} \left( \tilde{D}_{\rm b}^{\Delta \phi} \right) \right) \left[ -n : n, -n : n \right]$
  }
  $\widehat{\bm{\Theta}} = \argmax \left( \mathsf{R} \right)$

  \caption{\texttt{fastGlobalAlign}}
  \label{alg:correlation-search}
\end{algorithm}

\subsection{Automotive Radar Inverse Sensor Model}
\label{sec:inverse-sensor-model}

\begin{figure}[ht]
  \centering
  \includegraphics[width=\linewidth] {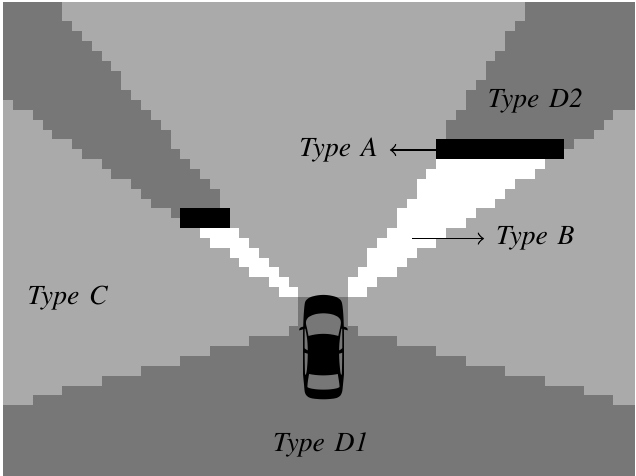}
  \caption{Schematic diagram showing four types of grid cells.}
  \label{fig:inverse-sensor}
\end{figure}

This section returns to the challenge of adapting the inverse sensor model
$p_{pq}(O~|~\bm{z}_k)$ to the measurement characteristics of automotive radar
sensors, introduced earlier in Sec.~\ref{sec:phd-approximation}.
Fig.~\ref{fig:inverse-sensor} shows a simplified radar scan $\bm{z}_k$ of an
underlying occupancy grid. For clarity of exposition, four distinct categories
of grid cells in Fig.~\ref{fig:inverse-sensor} are defined below:
\begin{itemize}
  \item \emph{Type A}: Grid cells in the vicinity of a radar range-azimuth
    return.
  \item \emph{Type B}: Grid cells along the path between the radar sensor and
    \emph{Type A} grid cells.
  \item \emph{Type C}: Grid cells in the ``viewshed'' of the radar sensor,
    i.e., within the radar field-of-view and not shadowed by another object,
    but not of \emph{Type A} or \emph{Type B}.
  \item \emph{Type D}: Grid cells outside the field-of-view of the radar
    (\emph{Type D1}) or shadowed by other objects closer to the radar
    (\emph{Type D2}).
\end{itemize}
The inverse sensor model must choose a $p_{pq}(O~|~\bm{z}_k)$ value for each of
these types of grid cells. In the following, the subscript $pq$ is dropped for
cleaner notation.

\subsubsection{Conventional Choices for the Inverse Sensor Model}

Since $\bm{z}_k$ provides no additional information on \emph{Type D} grid
cells, the occupancy in these cells is conditionally independent of $\bm{z}_k$,
that is
\[
  p^D(O~|~\bm{z}_k) = p(O)
\]
where $p(O)$ is the prior probability of occupancy defined earlier in
Sec.~\ref{sec:phd-localization}.

Grid cells of \emph{Type B} and \emph{Type C} may be hypothesized to have low
occupancy probability, since these grid cells were scanned by the sensor but no
return was obtained. As a result, conventionally
\[
  p^B(O~|~\bm{z}_k) \leq p(O)
\]
and
\[
  p^C(O~|~\bm{z}_k) \leq p(O)
\]

Finally, grid cells of \emph{Type A} may be hypothesized to have higher
occupancy probability, since a return has been observed in the vicinity of
these cells. Conventionally,
\[
  p^A(O~|~\bm{z}_k) \geq p(O)
\]
In the limit, if
the OGM grid cell size is comparable to the sensor range and angle uncertainty,
or if the number of scans is large enough such that the uncertainty is captured
empirically, only the grid cells that contain the sensor measurement may be
considered to be of \emph{Type A}.

\subsubsection{Automotive Radar Sensor Characteristics}

Intense clutter properties and sparsity of the automotive radar data complicate
the choice of the inverse sensor model.

\textbf{Sparsity.} First, sparsity of the radar scan implies that many occupied
\emph{Type A} grid cells in the radar environment might be incorrectly
categorized as free \emph{Type C} cells.  This can be observed in
Fig.~\ref{fig:radar-scan-matching}. As evidenced by the batch of scans in
Fig.~\ref{fig:radar-batch}, the radar environment is ``dense'' in that many
grid cells contain radar reflectors. However, any individual radar scan, such
as the one shown in Fig.~\ref{fig:single-scan}, suggests a much more sparse
radar environment. As a result, a grid cell which is occupied in truth will be
incorrectly categorized as \emph{Type C} in many scans, and correctly as
\emph{Type A} in a few scans.

The sparsity of radar returns also makes it challenging to distinguish
\emph{Type C} cells from cells of \emph{Type D2}. Since many occluding
obstacles are not detected in each scan, the occluded cells of \emph{Type D2}
are conflated with free cells of \emph{Type C}.

In context of the inverse sensor model, as the radar scan becomes more sparse
\[
  p^C(O~|~\bm{z}_k) \rightarrow {p^D(O~|~\bm{z}_k)}^{-}
\]
where the superscript $-$ denotes a limit approaching from below. Intuitively,
approaching $p^D(O~|~\bm{z}_k)$ implies that the measurement $\bm{z}_k$ is very
sparse in comparison to the true occupancy, and thus does not provide much
information on lack of occupancy.

\textbf{Clutter.} Second, there is the matter of clutter. The grid cells in the
vicinity of a clutter measurement may be incorrectly categorized as \emph{Type
A}, and the grid cells along the path between the radar and clutter measurement
may be incorrectly categorized as \emph{Type B}.

In context of the inverse sensor model, as the radar scan becomes more
cluttered
\begin{align*}
  p^B(O~|~\bm{z}_k) &\rightarrow {p^D(O~|~\bm{z}_k)}^{-} \\
  p^A(O~|~\bm{z}_k) &\rightarrow {p^D(O~|~\bm{z}_k)}^{+}
\end{align*}
where the superscript $+$ denotes a limit approaching from above.

\subsubsection{A Pessimistic Inverse Sensor Model}

The results presented in Sec.~\ref{sec:results} are based on a pessimistic
sensor model, such that $p^B(O~|~\bm{z}_k) = p^C(O~|~\bm{z}_k) =
p^D(O~|~\bm{z}_k)$. This model assumes that the radar measurements provide no
information about free space in the radar environment.

In particular, the inverse sensor model assumes
\[
  p^B(O~|~\bm{z}_k) = p^C(O~|~\bm{z}_k) = p^D(O~|~\bm{z}_k) = p(O) = 0.1
\]
and
\[
  p^A(O~|~\bm{z}_k) = 0.2
\]

\section{Experimental Results}
\label{sec:results}

The radar-based urban positioning system was evaluated experimentally using the
dataset described in~\cite{narula2020texcup}, collected on Thursday, May
9, 2019 and Sunday, May 12, 2019 during approximately \SI{1.5}{\hour} of
driving each day in and around the urban center of Austin, TX. This section
presents the evaluation results.

\subsection{Data Collection}

\begin{figure}[ht]
  \centering
  \includegraphics[width=\linewidth,trim={300 0 300 0},clip]{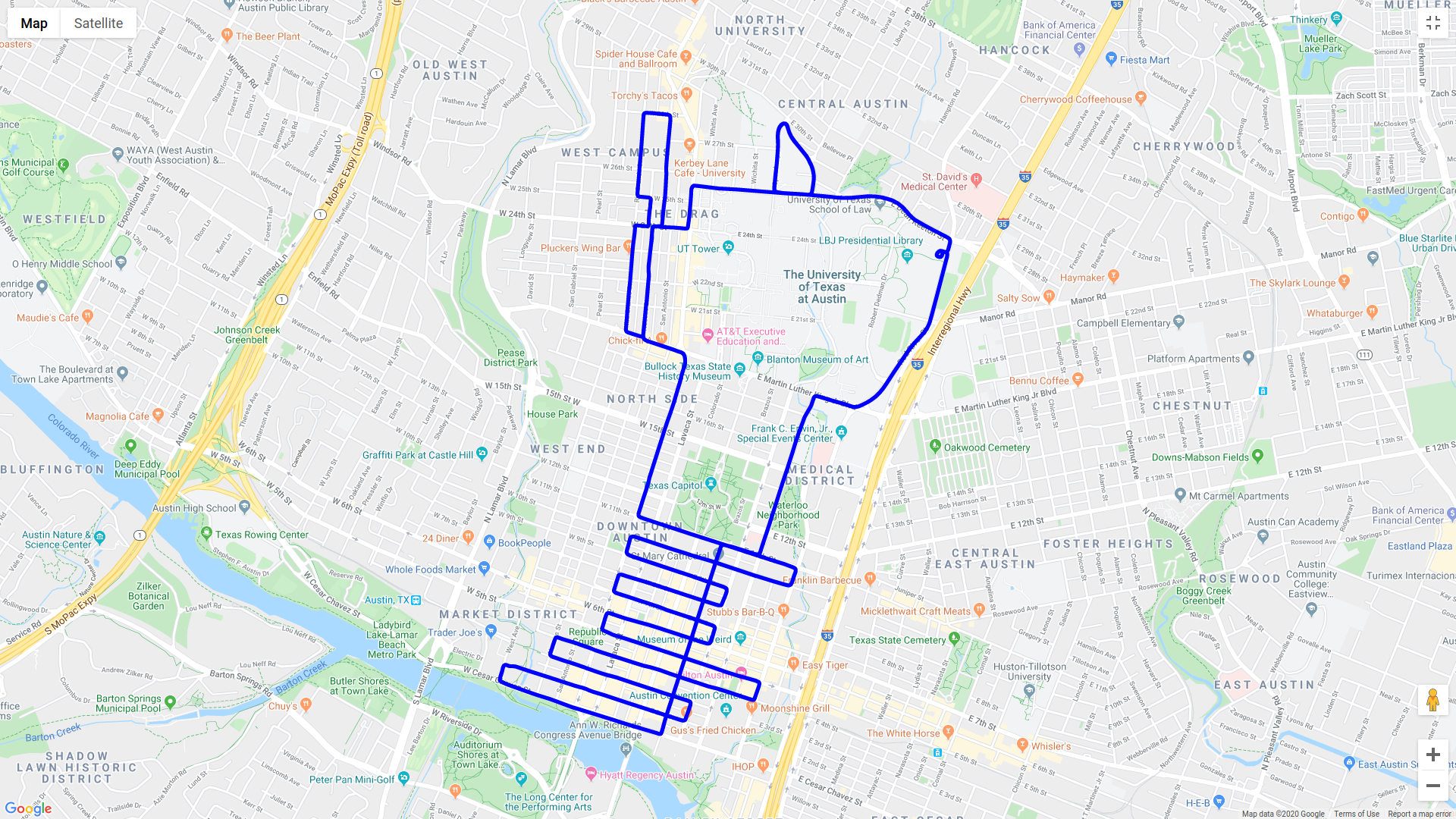}
  \caption{Test route through The University of Texas west campus and Austin
    downtown. These areas are the most challenging for precise GNSS-based
    positioning and thus would benefit the most from radar-based
    positioning. The route was driven once on a weekday and again on the
    weekend to evaluate robustness of the radar map to changes in traffic and
    parking patterns.}
  \label{fig:deep-urban-route}
\end{figure}

Fig.~\ref{fig:deep-urban-route} shows the route followed by the
radar-instrumented vehicle. Of particular interest is the southern part of the
route which combs through every street in the Austin, TX downtown area. Urban
canyons are the most challenging for precise GNSS-based
positioning~\cite{humphreys2019deepUrbanIts} and would benefit the most from
radar-based positioning.  The route was driven once on a weekday and again on
the weekend to evaluate robustness of the radar map to changes in traffic and
parking patterns.

\begin{figure}[ht]
  \centering
  \includegraphics[width=\linewidth,trim={50 100 50 125},clip]{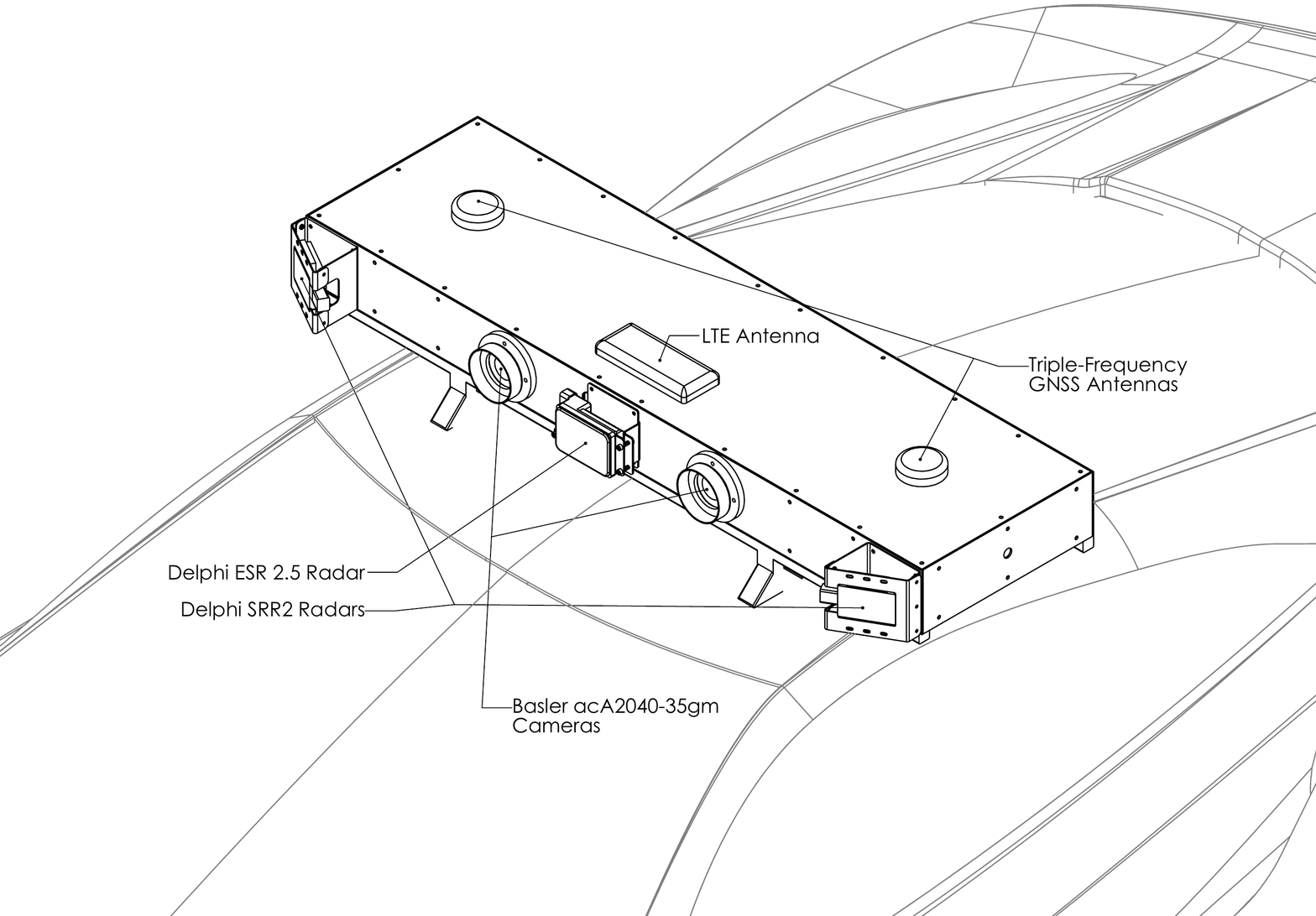}
  \caption{The University of Texas Sensorium is an integrated platform for
    automated and connected vehicle perception research.  It includes three
    automotive radar units, one electronically-scanning radar (ESR) and two
    short-range radars (SRR2s); stereo visible light cameras; automotive- and
    industrial-grade inertial measurement units (IMUs); a dual-antenna,
    multi-frequency software-defined GNSS receiver; 4G cellular connectivity;
    and a powerful internal computer. An iXblue ATLANS-C CDGNSS-disciplined INS
    (not shown) is mounted at the rear of the platform to provide the ground
    truth trajectory.}
  \label{fig:sensorium}
\end{figure}

The Delphi electronically-scanning radar (ESR) and short-range radars (SRR2s)
used in this study are mounted on an integrated perception platform called the
Sensorium, shown schematically in Fig.~\ref{fig:sensorium}. The coverage
patterns of the radar units are shown in Fig.~\ref{fig:radar-coverage}. The
Sensorium's onboard computer timestamps and logs the radar returns from the
three radar units.

\definecolor{SRR2}{HTML}{a8ddb5}
\definecolor{ESR}{HTML}{43a2ca}
\begin{figure}[ht]
  \centering
  \begin{tikzpicture}
    \draw[very thin, black] (-30mm,-12mm) rectangle (30mm,60mm);
    \node[] at (-28mm,-15mm) {\footnotesize \SI{-100}{\meter}};
    \node[] at (-36mm,-11mm) {\footnotesize \SI{-50}{\meter}};
    \node[] at (28mm,-15mm) {\footnotesize \SI{100}{\meter}};
    \node[] at (-36mm,59mm) {\footnotesize \SI{200}{\meter}};
    \fill[color=SRR2,opacity=0.25] (0,0) -- (23.18222mm,-6.211657mm) arc (-15:135:24mm) -- (0,0);
    \fill[color=SRR2,opacity=0.25] (0,0) -- (-23.18222mm,-6.211657mm) arc (195:45:24mm) -- (0,0);
    \fill[color=ESR,opacity=0.25] (0,0) -- (12.728mm,12.728mm) arc (45:135:18mm) -- (0,0);
    \fill[color=ESR,opacity=0.25] (0,0) -- (9.11653mm,51.7024mm) arc (80:100:52.5mm) -- (0,0);
    \node[inner sep=0pt] at (0,0) {\includegraphics[width=.15\linewidth,trim={0 100 0 0},clip]{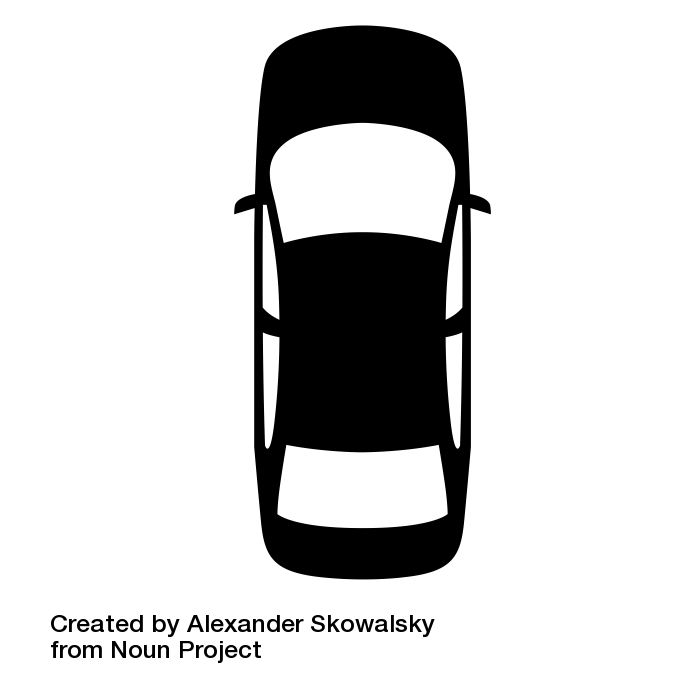}};
    \node[] at (-15mm,5mm) {\footnotesize SRR2 left};
    \node[] at (15mm,5mm) {\footnotesize SRR2 right};
    \node[] at (0mm,10mm) {\footnotesize ESR medium};
    \node[] at (0mm,40mm) {\footnotesize ESR long};
    \draw[very thin, black] (16.97mm,16.97mm) -- (22.62mm,22.62mm);
    \draw[very thin, black] (-16.97mm,16.97mm) -- (-22.62mm,22.62mm);
    \node[] at (-24mm,24mm) {\footnotesize $l_1$};
    \node[] at (24mm,24mm) {\footnotesize $l_2$};
  \end{tikzpicture}
  \caption{Coverage patterns for the three Sensorium radar units. The ESR
  provides simultaneous sensing in a narrow (\SI{+-10}{\degree}) long-range
  (\SI{175}{\meter}) coverage area and a wider (\SI{+-45}{\degree})
  medium-range (\SI{60}{\meter}) area. The SRR2 units each have a coverage area
  of \SI{+-75}{\degree} and \SI{80}{\meter}.  The line $l_1$ marks the
  left-most extent of the right SRR2's field of view.  Similarly, $l_2$ marks
  the right-most extent of the left SRR2's field of view.  Each SRR2 is
  installed facing outward from the centerline at an angle of
  \SI{30}{\degree}.}
  \label{fig:radar-coverage}
\end{figure}
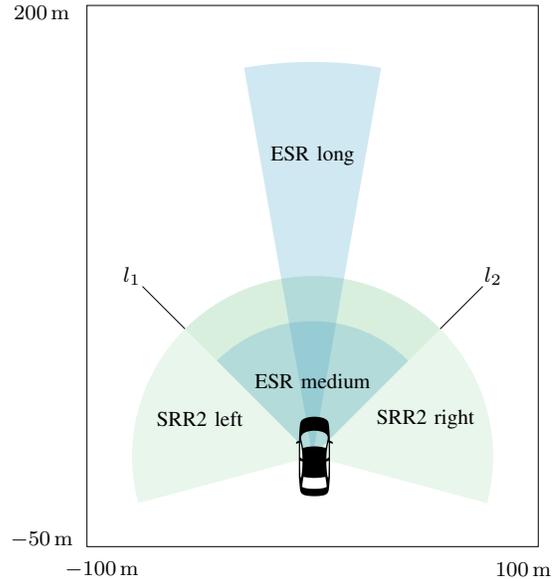

The reference position and orientation trajectory for the data are generated
with the iXblue ATLANS-C: a high-performance RTK-GNSS coupled fiber-optic
gyroscope (FOG) inertial navigation system (INS). The post-processed position
solution obtained from the ATLANS-C is decimeter-accurate throughout the
dataset.

\begin{figure*}[t]
  \centering
  \begin{minipage}[b]{0.325\textwidth}
    \centering
    \begin{tikzpicture}
      \node[inner sep=0pt] at (0,0)
      {\includegraphics[width=0.95\linewidth]{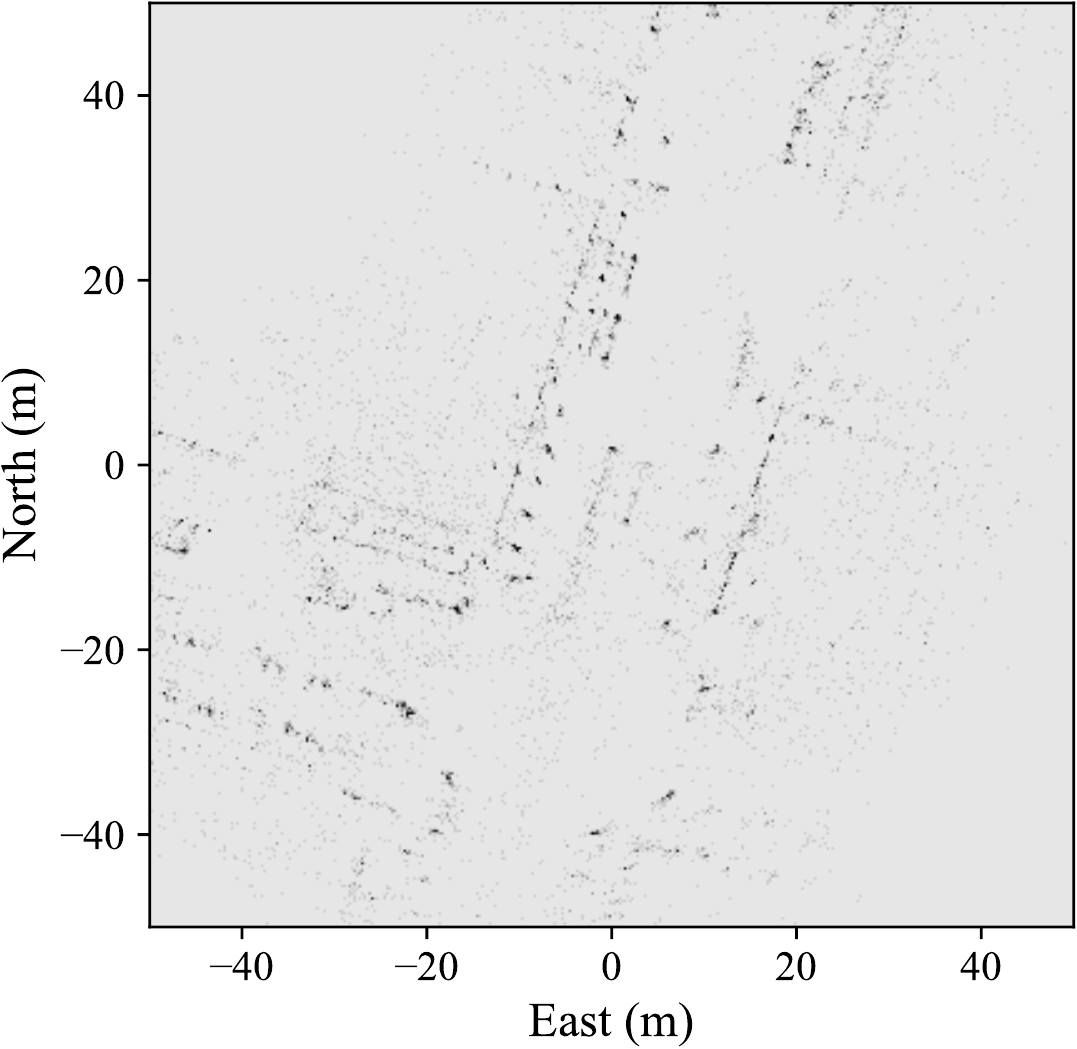}};
      \draw[color=red, rotate=-20] (-0.3,-0.6) rectangle +(0.5,2);
    \end{tikzpicture}
    \subcaption{}
    \label{fig:map-ogm}
  \end{minipage}
  \begin{minipage}[b]{0.325\textwidth}
    \centering
    \begin{tikzpicture}
      \node[inner sep=0pt] at (0,0)
      {\includegraphics[width=0.95\linewidth]{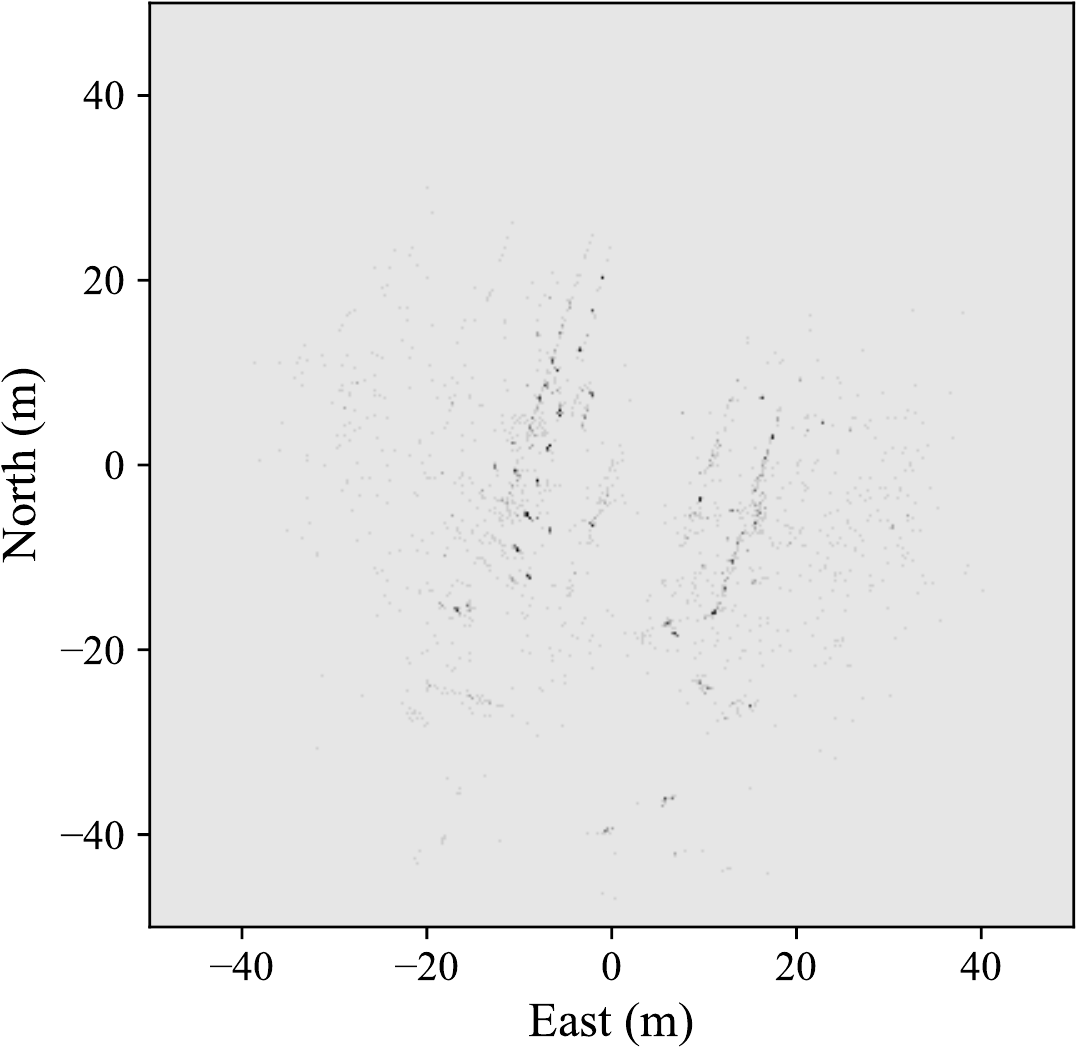}};
      \draw[color=red, rotate=-20] (-0.3,-0.6) rectangle +(0.5,2);
    \end{tikzpicture}
    \subcaption{}
    \label{fig:batch-ogm}
  \end{minipage}
  \begin{minipage}[b]{0.325\textwidth}
    \centering
    \begin{tikzpicture}
      \node[inner sep=0pt] at (0,0)
      {\includegraphics[width=0.95\linewidth]{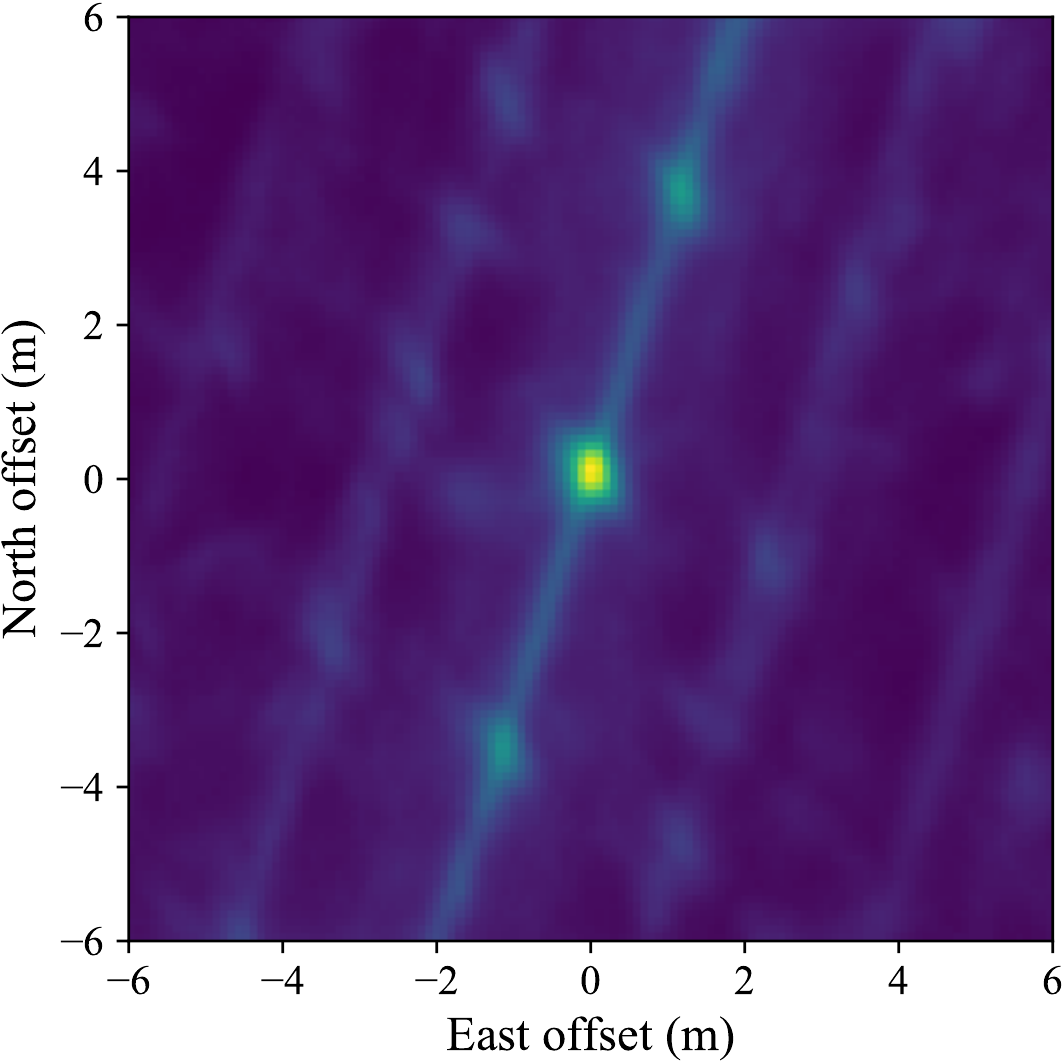}};
      \draw[color=red] (0.4,1.4) rectangle (1.2,2.2);
      \draw[color=red] (-0.5,-1.55) rectangle (0.3,-0.75);
    \end{tikzpicture}
    \subcaption{}
    \label{fig:xcorr}
  \end{minipage}
  \caption{This figure shows an interesting example of radar-based urban
  positioning with the proposed method. Panel (a) shows the occupancy grid
  estimated from the prior map point cloud. Panel (b) shows the same for a
  \SI{5}{\second} batch of scans collected in the same region.  For ease of
  visualization, the batch occupancy grid has already been aligned with the
  map occupancy grid.  Panel (c) shows the cross-correlation between the
  batch and map occupancy grids at $\Delta \phi =$ \SI{0}{\degree}.  Given
  that no rotational or translational offset error has been applied to the
  batch, the correlation peak should appear at $(0,0)$. The offset of the
  peak in panel (c) from $(0,0)$ is the translational estimate error of the
  proposed method. Also note the increased positioning uncertainty in the
  along-track direction, and the two local correlation peaks (marked with red
  squares in panel (c)) due to the repeating periodic pattern of radar
  reflectors in the map and the batch (marked with red rectangles in panels
  (a) and (b)).}
  \label{fig:map-batch-xcorr}
\end{figure*}
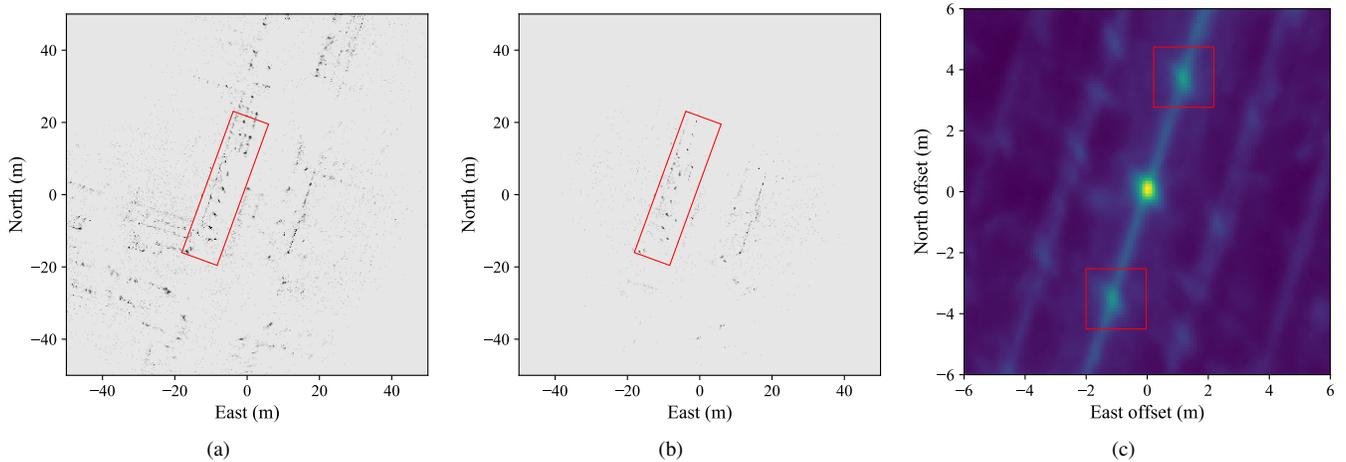

\subsection{Mapping}

The first step to radar-based localization is generation of a radar map point
cloud. Radar scans collected from the May 9, 2019 session are aggregated to
create a map with the benefit of the ATLANS-C reference trajectory. The map
point cloud is stored in a k-d tree for efficient querying during localization.

Two important notes are in order here. First, automotive radar clutter is
especially intense when the vehicle is stationary.  Accordingly, this paper
disregards radar returns obtained when the vehicle is moving slower than
\SI{1}{\meter\per\second} (approximately \num{2.23} miles per hour) for both
mapping and localization. This implies that radar-based localization is
only available during periods when the vehicle is moving faster than
\SI{1}{\meter\per\second}. Second, it was observed that radar returns far
from the vehicle are mostly clutter and have negligible resemblance to the
surrounding structure. As a result, this paper only considers radar returns
with range less than \SI{50}{\meter}. It must be noted that these two
parameters have not been optimized to produce the smallest estimation errors;
they have been fixed based on visual inspection.

\subsection{Localization Results with Perfect Odometry}

This section evaluates the localization performance of the proposed method on
the May 12, 2019 radar data for the case in which odometric drift over the radar
batch-of-scans interval is negligible.  With decreasing quality of the odometry
sensor(s), this assumption holds only over ever shorter batch intervals.
Therefore, the performance of the algorithm is evaluated for a range of batch
lengths.

\subsubsection{Test procedure}

A drift-free vehicle trajectory over a batch-of-scans interval is generated
with the reference solution from the iXblue ATLANS-C. This trajectory is then
artificially offset by a two-dimensional rigid transformation error. The
translational error is distributed such that
$\Delta \bm{t} \sim \mathcal{N}(\bm{0}, \sigma_t^2 I)$ with $\sigma_t =$
\SI{2}{\meter}, and the rotational error is distributed such that
$\Delta \phi \sim \mathcal{N}(0, \sigma_\phi^2)$ with $\sigma_\phi =$
\SI{3}{\degree}.  The proposed localization technique takes the
erroneously-offset position and heading trajectory as the initial guess of the
vehicle state.

The prior radar map point cloud in the vicinity of the initial guess of the
vehicle position is retrieved with a query to the k-d tree.  Additionally, the
batch of body-frame radar returns is transformed to a common reference frame
based on the erroneous trajectory.  The goal is to align the two point clouds
and thereby estimate the artificially-induced translational and rotational
offset.

As a first step, the map and batch occupancy grids are generated based on the
aggregated point clouds, following the procedure described in
Sec.~\ref{sec:technique} and~\ref{sec:implementation}. The extent of the
occupancy grids is determined by the bounds of the area scanned by the radars
during localization. Given the maximum range of radar returns considered here,
the correlation region is typically restricted to \SI{+-50}{\meter} around the
provided position at the end of the batch. With a grid cell size of
\SI{10}{\centi\meter}, the occupancy grid size in discrete coordinates is
typically on the order of $n = 1000$.  The translation search space is limited
to $\pm 3 \sigma_t = \pm$ \SI{6}{\meter}, resulting in $n_l = 120$. Similarly,
the rotation search space is limited to $\pm 3 \sigma_\phi = \pm$
\SI{9}{\degree} with \SI{1}{\degree} steps, resulting in $m = 18$.

\subsubsection{Five-second Batches}

This section evaluates and analyzes the proposed method for a fixed batch
length of \SI{5}{\second}. Fig.~\ref{fig:map-batch-xcorr} shows an
interesting example of radar-based urban positioning. For ease of visualization,
no translation or rotation offset error has been applied to the batch point
cloud. The occupancy grid estimated from the \SI{5}{\second} batch point
cloud is shown in Fig.~\ref{fig:batch-ogm}. Similarly,
Fig.~\ref{fig:map-ogm} shows the occupancy grid estimated from the map point
cloud retrieved from the map database. Fig.~\ref{fig:xcorr} shows the cross
correlation between the batch and map occupancy grids at $\Delta \phi =$
\SI{0}{\degree}. Given that no offset error has been applied to the batch,
one should expect the correlation peak to appear at $(0,0)$ in
Fig.~\ref{fig:xcorr}. The offset of the peak from $(0,0)$ in this case is the
translational estimate error.

\begin{figure}[ht]
  \centering
  \includegraphics[width=\linewidth] {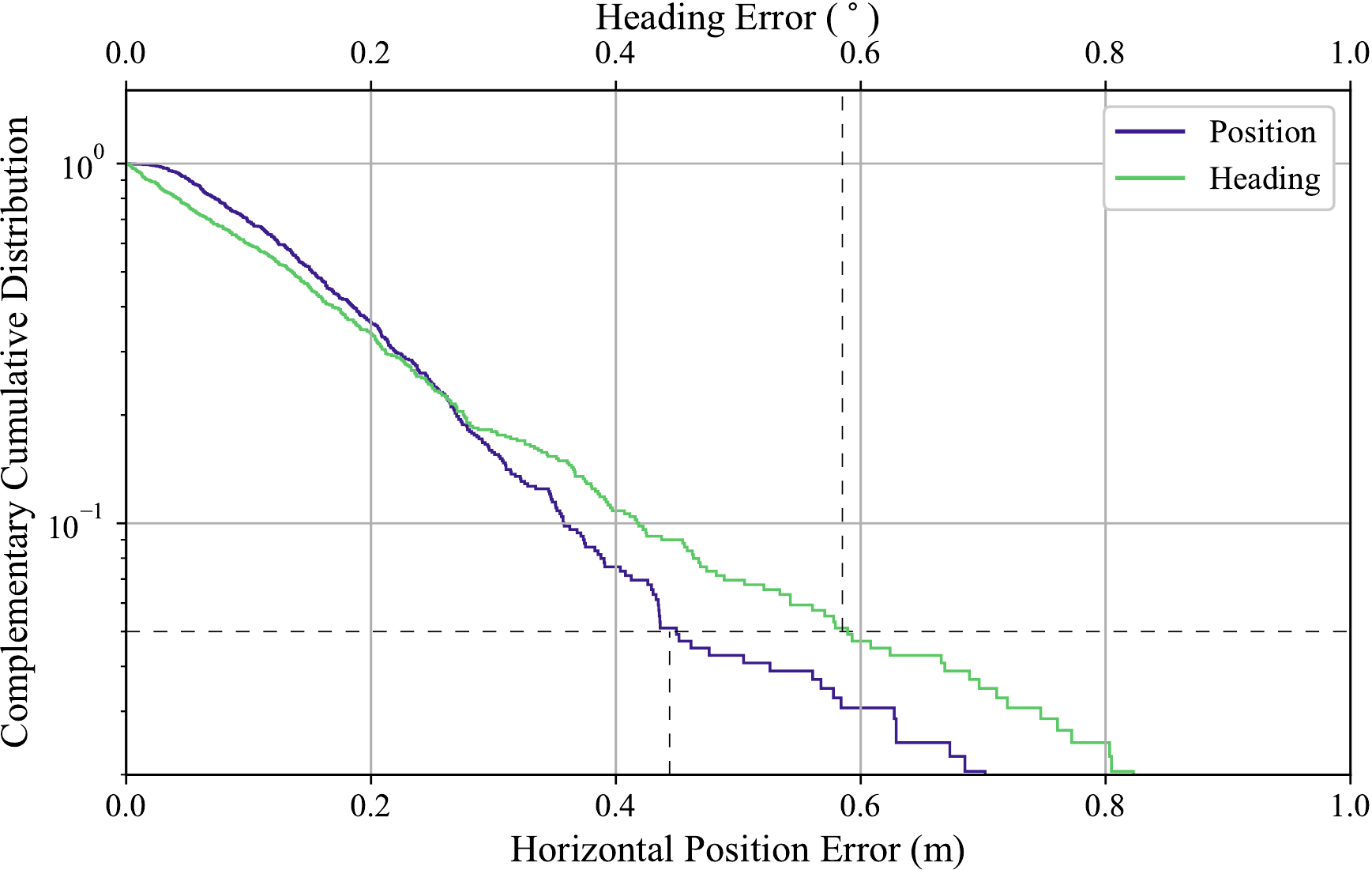}
  \caption{The complementary cumulative distribution (also known as a survival
  function) indicates how often (that is, in what fraction of \SI{5}{\second}
  epochs) the localization procedure in the text was found to exceed a given
  level of error.  The logarithmic vertical scale makes the tails of the
  distribution, corresponding to outliers that may cause tracking errors, more
  visible. For \SI{5}{\second} batches, the 95-percentile horizontal
  positioning error is observed to be \SI{44}{\centi\meter} and the
  95-percentile heading error is observed to be \SI{0.59}{\degree}.}
  \label{fig:ccdf-5sec}
\end{figure}

Two other interesting features of the cross-correlation in Fig.~\ref{fig:xcorr}
are worth noting. First, the correlation peak decays slower in the along-track
direction---in this case approximately aligned with the south-southwest
direction. This is a general feature observed throughout the dataset, since
most of the radar reflectors are aligned along the sides of the
streets. Second, there emerge two local correlation peaks offset by
$\approx$\SI{4}{\meter} along the direction of travel. These local peaks are
due to the repeating periodic structure of radar reflectors in both the map and
the batch occupancy grids. In other words, shifting the batch occupancy grid
forward or backward along the vehicle trajectory by $\approx$\SI{4}{\meter}
aligns the periodically-repeating reflectors in an off-by-one manner, leading
to another plausible solution.  Importantly, the uncertainty envelope of the
initial position peak can span several meters, encompassing both the global
optimum and one or more local optima.  This explains why gradient-based
methods, which seek the nearest optimum, are poorly suited for use with
automotive radar.

The complementary cumulative distribution functions (CCDF), i.e., the fraction
of epochs exceeding any given level of error, of the horizontal position and
heading error magnitudes are plotted on a log scale in Fig.~\ref{fig:ccdf-5sec}
for \SI{5}{\second} batches.  In 95\% of epochs, the horizontal position
error magnitude was no greater than \SI{44}{\centi\meter}, and the heading
error magnitude was no greater than \SI{0.59}{\degree}.

\subsubsection{Sensitivity to Batch Length}

The assumption of negligible odometric drift over the batch-of-scans interval
does not hold over \SI{5}{\second} for low-cost odometry sensors, but may hold
for longer than \SI{5}{\second} for high-performance sensors. Thus, is
important to evaluate the proposed localization technique's sensitivity to
batch length.

\begin{figure}[ht]
  \centering
  \includegraphics[width=\linewidth] {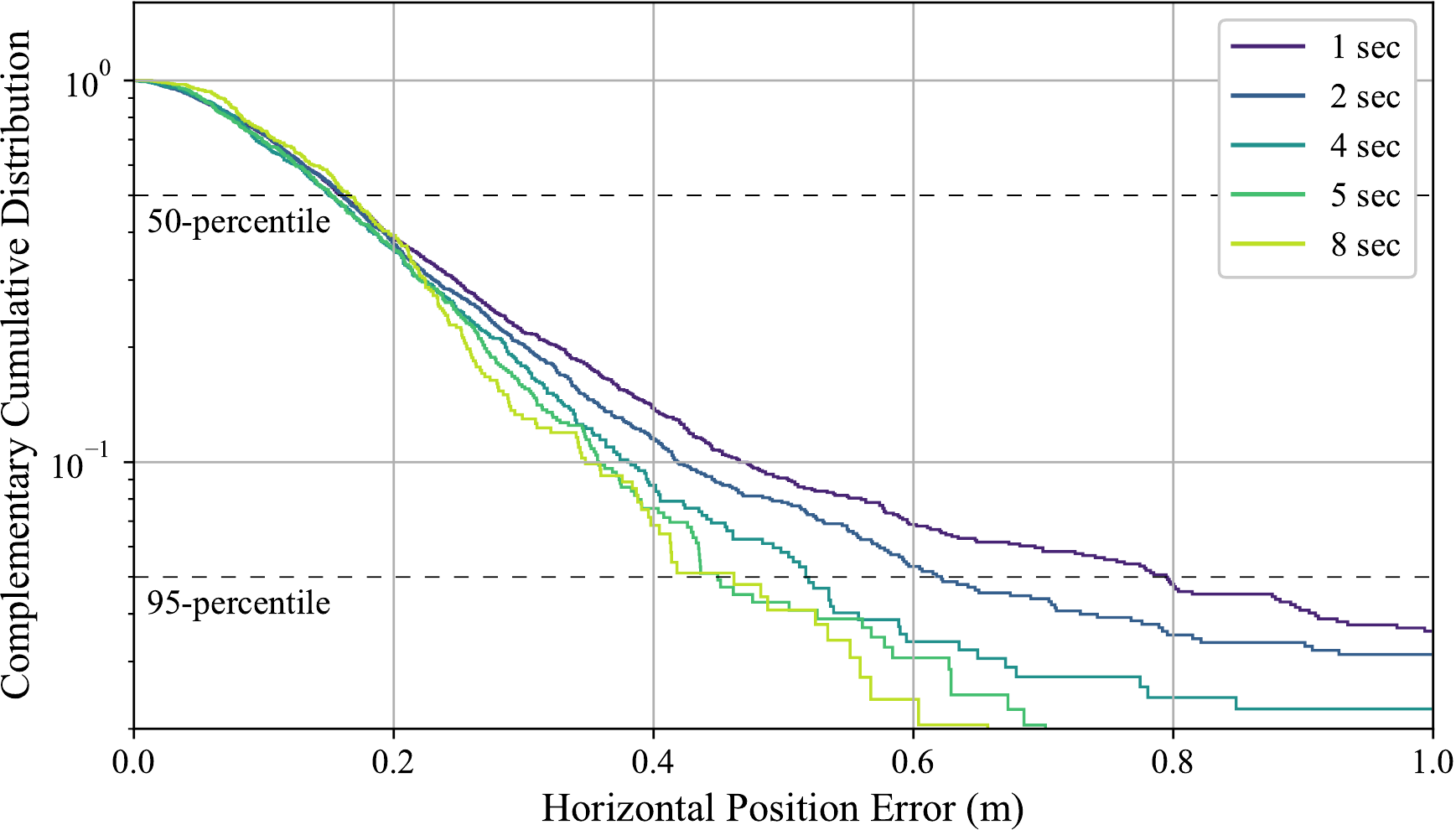}
  \caption{CCDFs for different batch lengths between \SI{1}{\second} and
  \SI{8}{\second}. The \num{50}-percentile errors are similar for shorter and
  longer batch lengths, but the difference becomes more noticeable at higher
  percentiles.}
  \label{fig:ccdf-fraglen}
\end{figure}

Fig.~\ref{fig:ccdf-fraglen} shows the CCDF for different batch lengths between
\SI{1}{\second} and \SI{8}{\second}. As expected, the errors are smaller for
longer batch lengths. It is interesting to note that the \num{50}-percentile
errors are similar for different batch lengths, but difference between the
CCDFs becomes more pronounced at higher percentiles.  This indicates that the
shorter batch lengths are adequate in most cases, but the longer batches help
contain the estimation error in a few cases.  Such behavior must be taken into
account in the integrity analysis of the system.  For example, while all batch
lengths exhibit similar \num{50}-percentile error behavior, batches shorter
than \SI{4}{\second} cannot be used in applications with a
\SI{50}{\centi\meter} alert limit and integrity risk smaller than \num{0.05}
per batch. On the other hand, in applications where \num{50}-percentile error
is the performance metric, it may be desirable to choose batches shorter than
\SI{4}{\second} to relax the requirements on short term odometric performance.

\subsection{Sensitivity to Odometric Drift}

Even over small batch durations, the odometric vehicle trajectory may exhibit
non-negligible drift. While the proposed technique does not provide a means to
estimate and eliminate such drift, it is important to study if such effects
lead to catastrophic degradation in performance.

This section considers simple drift models for position and heading drift
during batch generation. In particular, odometry-based orientation is
typically computed as integrated angular rate (e.g., from a gyroscope), and
odometry-based distance is typically derived as either integrated
speed (e.g., with rotary encoders, radar-based odometry), or as
double-integrated acceleration (e.g., with an accelerometer). To a coarse
approximation, this section assumes that the largest source of error in
odometry is due to integration of residual biases. This has the effect of
a linear drift in orientation estimates and either linear (in case of
integrated velocity measurements) or quadratic (in case of double-integrated
acceleration) drift in position estimates.

\begin{figure}[ht]
  \centering
  \includegraphics[width=\linewidth] {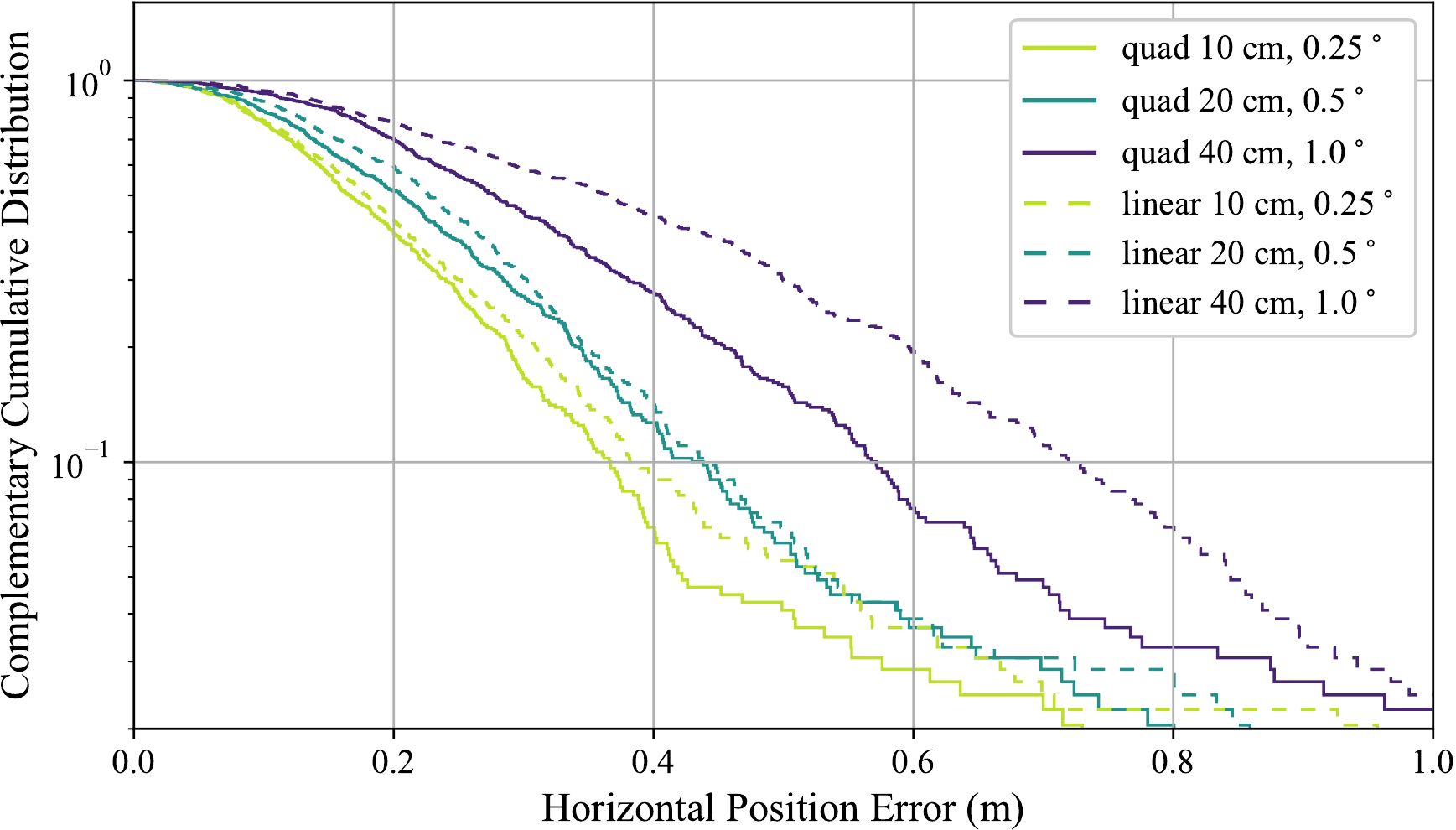}
  \caption{CCDFs for \SI{5}{\second} batches with three different levels of
  trajectory drift during batch generation. The solid lines represent
  quadratic drift in position and linear drift in heading. The dashed lines
  represent linear drift in both position and heading. The three different
  colors represent standard deviation of the drift at the end of the
  \SI{5}{\second} interval: (\SI{10}{\centi\meter}, \SI{0.25}{\degree}),
  (\SI{20}{\centi\meter}, \SI{0.5}{\degree}), and (\SI{40}{\centi\meter},
  \SI{1}{\degree}).}
  \label{fig:ccdf-drift}
\end{figure}

As before, a drift-free vehicle batch trajectory is obtained from the
reference solution from the iXblue ATLANS-C. In addition to the rigid
transformational offset error applied in the previous subsection, a
time-dependent position and heading drift is introduced in to the batch
trajectory. The trajectory is then used to transform the body-frame radar
returns in to a common reference frame as before. Assuming that the
odometric drift is small as compared to the rigid offset, the proposed approach
is evaluated in its ability to estimate the rigid offset.

This section restricts the analysis to \SI{5}{\second} batches with three
different levels of pose drift. The level of drift is characterized by the
standard deviation of the position and heading error at the end of the
\SI{5}{\second} interval. The three pairs of drift levels considered here are
chosen such that they are relatively small as compared to $\sigma_t$ and
$\sigma_\phi$: (\SI{10}{\centi\meter}, \SI{0.25}{\degree}),
(\SI{20}{\centi\meter}, \SI{0.5}{\degree}), and (\SI{40}{\centi\meter},
\SI{1}{\degree}). Both the linear and quadratic error growth models for
position drift are explored.

Fig.~\ref{fig:ccdf-drift} shows the CCDF plots for different levels of
odometric drift and different position drift models. As expected, the
performance degrades with larger drift during batch generation.
Nevertheless, the effect of impairment is not catastrophic---with quadratic
position and linear heading drift standard deviation of \SI{40}{\centi\meter}
and \SI{1}{\degree} at \SI{5}{\second}, respectively, the \num{95}-percentile
error in estimation of the rigid offset is \SI{67}{\centi\meter} and
\ang{1.17}.

\section{Conclusion}
\label{sec:conclusion}

A novel technique for robust 50-cm-accurate urban ground positioning based on
commercially-available low-cost automotive radars has been proposed. This is a
significant development in the field of AGV localization, which has
traditionally been based on sensors such as lidar and cameras that perform
poorly in bad weather conditions. The proposed technique is computationally
efficient yet obtains a globally-optimal translation and heading solution,
avoiding local minima caused by repeating patterns in the urban radar
environment.  Performance evaluation on an extensive and realistic urban data
set shows that, when coupled with stable short-term odometry, the technique
maintains \num{95}-percentile errors below \SI{50}{\centi\meter} in horizontal
position and \ang{1} in heading.

\section*{Acknowledgments}
This work has been supported by Honda R\&D Americas through The University of
Texas Situation-Aware Vehicular Engineering Systems (SAVES) Center
(http://utsaves.org/), an initiative of the UT Wireless Networking and
Communications Group; by the National Science Foundation under Grant
No. 1454474 (CAREER); and by the Data-supported Transportation Operations and
Planning Center (DSTOP), a Tier 1 USDOT University Transportation Center.

\bibliographystyle{IEEEtran} 
\bibliography{pangea}

\balance

\end{document}